\documentclass[aps,prl,reprint,superscriptaddress,nofootinbib]{revtex4-2}

\usepackage[T1]{fontenc}
\usepackage{lmodern}
\usepackage{amsmath,amssymb,amsfonts,amsthm,mathtools,bm}
\usepackage[notrig]{physics}
\usepackage{etoolbox}
\usepackage{graphicx}
\usepackage{dsfont}
\usepackage[hidelinks]{hyperref}

\newcommand{\dvol}{\operatorname{dvol}}
\newcommand{\dvolg}{\operatorname{dvol}_{g_t}}

\renewcommand{\dd}{\mathrm{d}}

\newcommand{\B}{\mathrm{B}}

\newcommand{\pd}{\partial}
\newcommand{\nab}{\nabla}

\newcommand{\kB}{k_\mathrm{B}}

\allowdisplaybreaks[1]

\newcommand{\parhead}[1]{\par\smallskip\noindent\textit{#1--- }}

\makeatletter
\auto@bib@empty
\let\auto@bib@innerbib\@empty
\let\@bibdataout@init\relax
\let\write@bibliographystyle\relax
\makeatother

\begin{document}

\title{Stochastic Thermodynamics on Time-Evolving Curved Spaces}

\author{Rihito Nagase}
\affiliation{Department of Applied Physics, The University of Tokyo, 7-3-1 Hongo, Bunkyo-ku, Tokyo 113-8656, Japan}

\author{Shoki Sugimoto}
\affiliation{Department of Applied Physics, The University of Tokyo, 7-3-1 Hongo, Bunkyo-ku, Tokyo 113-8656, Japan}
\affiliation{{Nonequilibrium Quantum Statistical Mechanics RIKEN Hakubi Research Team, RIKEN Pioneering Research Institute (PRI), Wako, Saitama 351-0198, Japan}}

\author{Asuka Takatsu}
\affiliation{Graduate School of Mathematical Sciences, The University of Tokyo, 3-8-1 Komaba, Meguro-ku, Tokyo 153-8914, Japan}
\affiliation{RIKEN Center for Advanced Intelligence Project (AIP), 1-4-1 Nihonbashi, Chuo-ku, Tokyo 103-0027, Japan}

\author{Takahiro Sagawa}
\affiliation{Department of Applied Physics, The University of Tokyo, 7-3-1 Hongo, Bunkyo-ku, Tokyo 113-8656, Japan}
\affiliation{Quantum-Phase Electronics Center (QPEC), The University of Tokyo, 7-3-1 Hongo, Bunkyo-ku, Tokyo 113-8656, Japan}
\affiliation{Inamori Research Institute for Science (InaRIS), Kyoto-shi, Kyoto 600-8411, Japan}

\begin{abstract}
We construct stochastic thermodynamics of overdamped Langevin systems on {nonrelativistic} curved spaces with time-dependent metrics. 
{
We identify a geometric contribution to heat and work arising from the time dependence of the metric, derived from the first law of thermodynamics and the definition of entropy production consistent with the second law.
The correction can also be interpreted as work done by the time-dependent metric on the kinetic energy, which is instantaneously dissipated as heat in the overdamped regime. 
}
As a special case, when the metric evolves according to {backward} Ricci flow, the entropy balance exhibits a structure similar to Perelman's {entropy} functional. {Our framework provides a way to quantify} thermodynamic costs in dynamics on time-evolving spaces such as diffusion on membranes.\end{abstract}

\maketitle

\parhead{Introduction}Stochastic thermodynamics is a framework for identifying thermodynamic costs in nonequilibrium processes~\cite{Sekimoto_1998,Crooks_1999,Seifert2012,Jarzynski_2011,Bustamante_2005,Ciliberto2017}. While originally formulated on {Euclidean spaces}, it has been extended to {curved spaces modeled by Riemannian manifolds} to describe a broader class of {stochastic processes}. {Whereas} a notable example is relativistic thermodynamics~\cite{Debbasch_2004,DunkelHanggi2005a,DunkelHanggi2005b,Fingerle_2007,Pal_2020,Cai2023Relativistic,Cai2023RelativisticII,CaiWangZhao2024FluctuationTheoremRM,Cai_2025}, nonrelativistic covariant thermodynamics~\cite{Graham1977,vanKampen1986,Polettini2013,Ding2020PRR,Ding2022PRR,Diosi2024RMP,Nakamura_2024,Miller2025PRSA} has also been developed for describing nonrelativistic phenomena on curved spaces~\cite{N_meth_2025,Reister_2005,Raible_2004,LittlejohnReinsch1997}, including particle motion on {surfaces}~\cite{Marque2025}, diffusion on cell membranes~\cite{Metzler_2016}, and Brownian motion of rotating magnetic particles~\cite{Romodina_2016}.\par
\begin{figure}[t]
  \centering
  \includegraphics[width=\columnwidth]{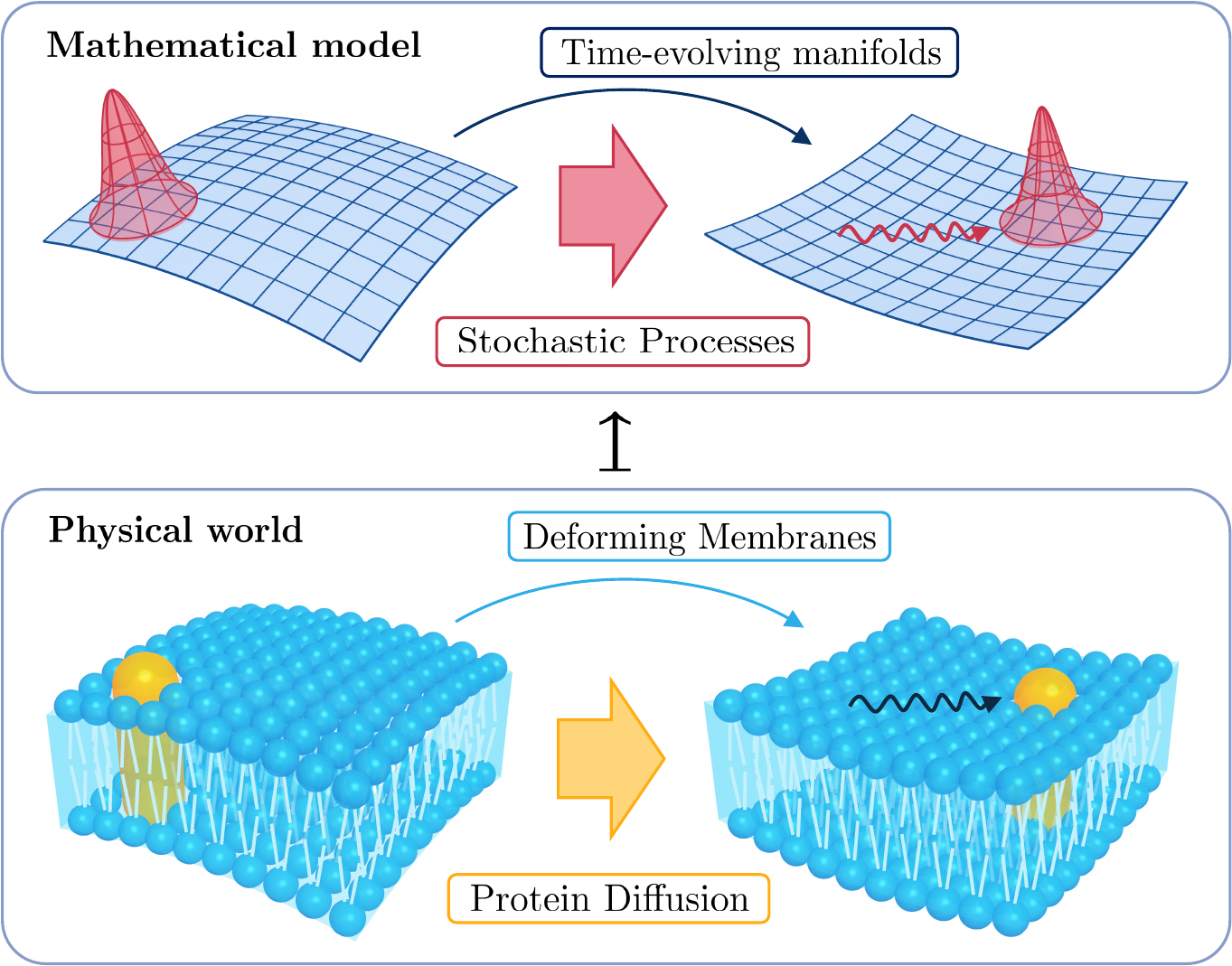}
  \caption{
  Correspondence between the mathematical model and protein diffusion on a biological membrane. The diffusing protein (yellow) is on the biological membrane (cyan) which generally deforms in time. Correspondingly, when modeled as a stochastic process, the manifold (blue) on which the distribution (red) is defined becomes time-dependent.
  }
  \label{fig:manifold}
\end{figure}

In many of these studies, the background space is assumed to be static. However, time-evolving background spaces also constitute an important class to be considered. {In mathematics}, such cases have been studied extensively in the context of Ricci flow, including studies on Brownian motion~\cite{CoulibalyPasquier2011}, entropy and optimal transport~\cite{Lott2009OptimalTransport,McCannTopping2010,Topping2009LOptimal}, stochastic analysis on path spaces~\cite{HaslhoferNaber2018,Kennedy2023} and {the geometrization conjecture which includes the Poincar\'{e} conjecture~\cite{perelman2002entropy,perelman2003ricciflowsurgerythreemanifolds,perelman2003finiteextinctiontimesolutions}}. 
{In physics}, various phenomena such as colloidal dynamics on time-evolving interfaces~\cite{Marque2025}, protein diffusion on fluctuating membranes~\cite{Quemeneur_2014}, Turing patterns on growing domains~\cite{Konow2019}, and chemical reaction on dynamically evolving electrode surfaces~\cite{Wood2016} are described as dynamics on time-evolving manifolds~\cite{Naji_2007,Duncan_2015,H_kansson_2018,bell2026surface,Krause2019,Bozzini2025} (Fig.~\ref{fig:manifold}). 
Yet how heat and work are expressed on time-evolving spaces, and how the fundamental laws of thermodynamics are formulated in general, remain to be understood.
\par
In this Letter, we consider an overdamped Langevin system on a time-evolving {nonrelativistic} curved space. We model the time-evolving space as a {manifold equipped with a time-dependent Riemannian metric} and formulate stochastic thermodynamics on it. The central claim is the identification of heat and work in a thermodynamically consistent manner. The time dependence of the metric contributes to both heat and work: {the metric} performs work on the kinetic energy, which is instantaneously dissipated as heat in the overdamped regime. This is a nontrivial energy balance that cannot be captured by {naively} decomposing the time derivative of the energy. We show that those correction terms to heat and work are consistent with the second law of thermodynamics with an appropriate definition of entropy production. 
{
Since our framework applies to dynamics on manifolds of arbitrary dimension, it covers dynamics in general configuration spaces beyond subspaces embedded in $\mathbb{R}^3$, such as systems with rotational degrees of freedom~\cite{N_meth_2025} and many-body systems with internal degrees of freedom~\cite{LittlejohnReinsch1997}.}
As an example, we present the explicit correction term for {a colloidal chain with a time-dependent end-to-end distance}. {Moreover}, we consider the Riemannian metric evolving under {backward Ricci flow} and show that the entropy balance has a structure analogous to Perelman's {entropy functional called $\mathcal{F}$-functional}.
\parhead{Setup}Consider {a system with a $d$-dimensional configuration space $M$,} coupled to a work source and a heat bath. The heat bath is assumed to {thermalize} much faster than the {system's} dynamics and the time evolution of the manifold, and to be in equilibrium at temperature $T$. 
The {system} is in general in a nonequilibrium state, and we consider its Langevin dynamics in the overdamped regime. 
The time evolution of {the space and the relative motion of the bath are} encoded in a time-dependent metric $g_t${, as discussed in Sec.~\ref{sec:supp_embedded} of the Supplemental Materials (SM)}. We fix the units for lengths and volumes throughout, and treat dimensionless quantities based on these units.
\par
We here summarize our notation. We denote the {state of the system} by $x\in M$ and take {a time-independent 
coordinate system} $(x^\mu)$. {Throughout, we employ Einstein's summation convention, whereby repeated upper and lower indices imply summation over the corresponding index.} Covariant and contravariant components are indicated by lower and upper indices, respectively. {The metric tensor} is represented by the covariant tensor $(g_t)_{\mu\nu}$, which defines the 
line element $\dd s^2=(g_t)_{\mu\nu}\dd x^\mu\dd x^\nu$. The inverse metric $(g_t)^{\mu\nu}$ is defined by $(g_t)^{\mu\kappa}(g_t)_{\kappa\nu}=\delta^\mu{}_\nu$ where $\delta^\mu{}_\nu$ is the Kronecker delta. {We denote by $\nabla_\mu$ the covariant derivative of $g_t$. For any scalar function $\phi$, we have $\nabla_\mu \phi = \partial_\mu \phi$.}
\par
We quantify the forces on the {system} from the work source and the heat bath. The work source applies the conservative force $(f_t)_\mu\coloneqq-\nabla_\mu\phi_t$ from the potential $\phi_t(x)$. For simplicity, we assume the absence of nonconservative forces. {The heat bath induces dissipation and fluctuations. 
We {consider generally anisotropic dissipation with a symmetric and positive-definite friction tensor $(\gamma_t)_{\mu\nu}$.} By assuming the fluctuation--dissipation relation to ensure relaxation of the momentum to the Maxwell distribution in the overdamped regime, we obtain the diffusion tensor {$(D_t)^{\mu\nu}=k_{\rm B}T(\gamma_t^{-1})^{\mu\nu}$, with $\kB$ being the Boltzmann constant and $(\gamma_t^{-1})^{\lambda\nu}$ being the inverse friction tensor satisfying $(\gamma_t)_{\mu\lambda}(\gamma_t^{-1})^{\lambda\nu}=\delta_\mu{}^\nu$. 
}}\par
Let $\dd\mu_t(x)$ be the probability measure for the {system} to be {in state} $x$ at time $t$, and denote its density function with respect to the volume measure $\dvolg=\sqrt{\det [(g_t)_{\mu\nu}]}\,\dd x$ by
\begin{align}
  \rho_t(x)\coloneqq\frac{\dd\mu_t(x)}{\dvolg}.\label{eq:def_rho}
\end{align}
We note that $\rho_t$ is independent of the choice of coordinate system $(x^\mu)$, since $\rho_t$ is the density with respect to the coordinate-independent volume measure $\mathrm{vol}_{g_t}$ (see {SM} for details). {Then}, $\rho_t$ obeys the overdamped Fokker--Planck {equation}
\begin{align}
  &\partial_t\left(\rho_t\dvolg\right)=-[\nabla_\mu (J_t)^\mu]\dvolg,
  \label{eq:conservation}\\
  &(J_t)^\mu\coloneqq-(D_t)^{\mu\nu}\left(\nabla_{\nu}\rho_t-\frac{\rho_t}{k_\B T}(f_t)_\nu\right).\label{eq:def_J_setup}
\end{align}
{This equation} is derived by imposing probability conservation 
$\frac{\dd}{\dd t}\int_U\rho_t\dvol_{g_t}=-\int_{\partial U}(J_t)^\mu n_\mu\dd A_{g_t}$ 
for $U\subset M$, with $n_\mu$ the covariant components of the outward unit normal to the boundary $\partial U$ and $\dd A_{g_t}$ {the Riemannian volume measure on $\partial U$} induced by $g_t$.
An equivalent form of {this Fokker--Planck equation} has also been studied in mathematics~\cite{Lenz_2011,Dziuk_2013}. We also note that it reduces to the overdamped Fokker--Planck equation on a static curved space in Refs.~\cite{Polettini2013,Diosi2024RMP} when $g_t$ is time-independent.
At the instantaneous equilibrium distribution $\rho_t^\mathrm{eq}(x)\propto\exp[-\phi_t(x)/(k_\B T)]$, we have $(J_t)^\mu=0$. 
\parhead{Main claim}Based on the above dynamics, we formulate stochastic thermodynamics in a thermodynamically consistent manner.
The main claim of this Letter is that {the heat rate $\dot{Q}_t$ absorbed by the system and the work rate $\dot{W}_t$ done on the system} are 
identified as
\begin{align}
\dot{Q}_t
&=\int_M
(J_t)^\mu[\nabla_\mu\phi_t]\dvolg
-\dot{G}_t,
\label{eq:Q_od_rate}\\
\dot{W}_t
&=
\int_M\rho_t(\pd_t\phi_t)\dvolg+\dot{G}_t.
\label{eq:W_od_rate}
\end{align}
Here $\dot{G}_t$ is {the} correction arising from the time dependence of the metric, defined by
\begin{align}
    \dot{G}_t\coloneqq-\frac{k_\B T}{2}\int_M\rho_t\operatorname{tr}_{g_t}[\partial_tg_t]\dvolg,
\label{eq:G_od_rate}
\end{align}
where $\operatorname{tr}_{g_t}[\partial_tg_t]\coloneqq(g_t)^{\mu\nu}\partial_t(g_t)_{\mu\nu}$.
When $g_t$ is time-independent, $\dot{G}_t=0$ and we recover the definitions of heat and 
work in Refs.~\cite{Polettini2013,Diosi2024RMP}. Since $\rho_t$, $\operatorname{tr}_{g_t}[\partial_tg]$, and $\dvol_{g_t}$ are all 
{coordinate free}, $\dot{G}_t$ is also invariant under coordinate transformations.\par
{We now derive the expressions for heat (\ref{eq:Q_od_rate}) and work (\ref{eq:W_od_rate}) from thermodynamic consistency.}
{For that purpose}, we define entropy production via {the rate of decrease of the relative entropy with respect to the instantaneous equilibrium measure}. {This, along with the first law of thermodynamics, will lead to the heat expression~(\ref{eq:Q_od_rate}) and the work expression~(\ref{eq:W_od_rate})}.\par
We first define entropy production.
\begingroup
With the instantaneous equilibrium measure
$\dd{\mu_t^{\rm eq}}(x)=\rho_t^{\mathrm{eq}}(x)\dvol_{g_t}$, we define the entropy
production rate $\dot{\Sigma}_t$ as
\begin{align}
  \dot{\Sigma}_t
  \coloneqq-\left.\frac{\dd}{\dd {\tau}}S(\mu_{\tau}||{\mu_t^{\rm eq}})\right|_{{\tau}=t},
  \label{eq:sigma_od_relent}
\end{align}
\endgroup
where the relative entropy between measures $\nu$ and $\nu'$ on a space $A$ is defined
as $S(\nu||\nu')\coloneqq\int_A\ln(\dd\nu/\dd\nu')\dd\nu$.
{Here $\mu_\tau$ follows Eq.~(\ref{eq:conservation}) in the derivative.}
This is a generalization of a result known for flat spaces~\cite{Van_den_Broeck_2010} to time-evolving curved spaces. {This definition can also be formulated in terms of the
forward and backward processes (see {SM} for details).}\par
{We obtain another expression for the entropy production rate by writing the probability current as}
\begin{align}
    (J_t)^\mu=-\rho_t (D_t)^{\mu\nu}\nabla_\nu\ln\frac{\rho_t}{\rho_t^\mathrm{eq}},\label{eq:J_grad}
\end{align}
which reflects the fact that {Eq.~(\ref{eq:conservation})} is a gradient flow of the relative 
entropy~\cite{JordanKinderlehrerOtto1998,Lisini_2008,Erbar_2010} (see {SM} for details). Evaluating the right-hand side of Eq.~(\ref{eq:sigma_od_relent}) using this expression, and defining the norm of a {covariant vector $V$ with respect to a positive-definite symmetric tensor $A$ as ${\|V\|_{A}}^2\coloneqq A^{\mu\nu}V_\mu V_\nu$, }we obtain
\begin{align}
    \dot{\Sigma}_t&=-\int_M\ln\frac{\dd\mu_t}{\dd{\mu_t^{\rm eq}}}\partial_t\dd\mu_t\\
    &=\int_M\left\|\nab\ln\frac{\rho_t}{\rho_t^\mathrm{eq}}\right\|_{D_t}^2\rho_t\dvolg,\label{eq:diff_RelativeEntropy}
\end{align}
{where $\nabla$ represents the differential.} The second law $\dot\Sigma_t\geq0$ follows immediately, {with equality achieved {if} $\rho_t=\rho_t^\mathrm{eq}$}.\par
We next {discuss} the heat expression~(\ref{eq:Q_od_rate}).
{We identify the entropy production as the total entropy change of the system and the 
heat bath},
\begin{align}
  \dot{\Sigma}_t
  =
  \dot{S}_t^\mathrm{sys}+\dot{S}_t^\mathrm{bath},\label{eq:EP_total}
\end{align}
{where $\dot{S}_t^\mathrm{bath}=-(\kB T)^{-1}\dot{Q}_t$}. Heat is thus determined by
\begin{align}
    \dot{Q}_t=-\kB T\left(\dot{\Sigma}_t-\dot{S}_t^\mathrm{sys}\right).\label{eq:def_heat}
\end{align}
The system entropy is given by the negative relative entropy with respect to the volume 
measure $\mathrm{vol}_{g_t}$ as~\cite{Miller2025PRSA}
\begin{align}
    S_t^\mathrm{sys}\coloneqq -S(\mu_t||\mathrm{vol}_{g_t})=-\int_M \rho_t\ln\rho_t\,\dd{\rm vol}_{g_t}.\label{eq:def_S_sys}
\end{align}
Applying {Eq.~(\ref{eq:conservation})} and integrating by parts to compute its time derivative, we obtain
\begin{align}
    \dot{S}_t^\mathrm{sys}
    &=-\int_M(\ln\rho_t+1)\,\partial_t(\rho_t\dvol_{g_t})+\int_M\rho_t\,\partial_t\dvol_{g_t}\\
    &=-\int_M(J_t)^\mu[\nabla_\mu \ln\rho_t]\dvolg+\int_M\rho_t\,\partial_t\dvolg.\label{eq:diff_S_sys}
\end{align}
Multiplying by $k_\B T$ and substituting 
$\partial_t\dvol_{g_t}=\frac{1}{2}\operatorname{tr}_{g_t}[\partial_tg]\dvol_{g_t}$, 
the second term yields $-\dot{G}_t$. Substituting this together with Eq.~(\ref{eq:diff_RelativeEntropy}) into Eq.~(\ref{eq:def_heat}) yields the heat expression~(\ref{eq:Q_od_rate}).\par
We next {discuss} the work expression~(\ref{eq:W_od_rate}) from the first law.
The energy function in the overdamped regime is given by
\begin{align}
  E_t\coloneqq \frac{d}{2}\kB T+\int_M\phi_t\rho_t\,\dvolg.\label{eq:energy_od}
\end{align}
Applying {Eq.~(\ref{eq:conservation})} and integration by parts, we obtain
\begin{align}
    \dot{E}_t
    &=\int_M\phi_t\,\partial_t(\rho_t\dvol_{g_t})+\int_M\rho_t(\partial_t\phi_t)\dvolg\\
  &=\int_M(J_t)^\mu[\nabla_\mu \phi_t]\dvolg+\int_M\rho_t(\partial_t\phi_t)\dvolg.\label{eq:1st_law_od_pre}
\end{align}
{The first law requires the decomposition}
\begin{align}
  \dot{E}_t
  &=\dot{Q}_t+\dot{W}_t.\label{eq:1st_law_od}
\end{align}
Substituting the heat expression~(\ref{eq:Q_od_rate}) into this equation yields the work 
expression~(\ref{eq:W_od_rate}).\par
\begin{figure}[t]
  \centering
  \includegraphics[width=\columnwidth]{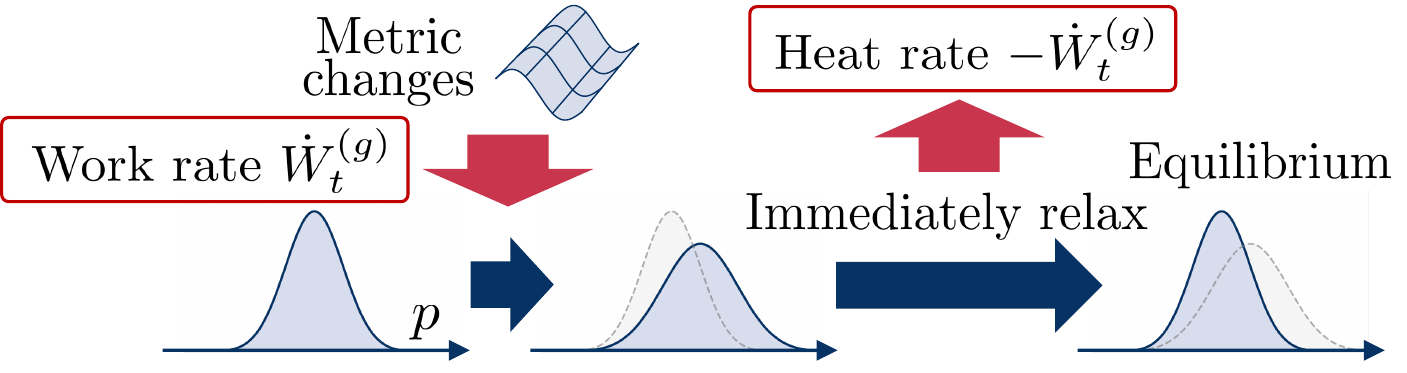}
  \caption{
  Schematic of the energy balance. The time-dependent metric tensor {performs work on the kinetic energy}, which is instantaneously dissipated as heat through momentum relaxation {in the overdamped regime}.
  }
  \label{fig:thermo}
\end{figure}
\par
{\parhead{Physical interpretation}We next give a physical interpretation of the geometric correction $\dot{G}_t$; it represents the work done by the time-dependent metric on the kinetic energy, which is instantaneously dissipated as heat in the overdamped regime (Fig.~\ref{fig:thermo}).}
{Let {$\pi_\mu\coloneqq m(g_t)_{\mu\nu}\dot{x}^{\nu}$} denote the covariant {component} of the {kinetic momentum} of {the degree of freedom $x^\mu$, where $m$ is a fixed mass scale}.}
{Since the transformation from $\pi_\mu$ to the conjugate momentum has a unit Jacobian, Liouville's theorem on a curved space preserves the measure on $(x^\mu,\pi_\mu)$~\cite{Marsden_1999}.}
Note that, in stochastic thermodynamics, it is standard to define heat and work with respect to the variables satisfying Liouville's theorem~\cite{Jarzynski_1997,Jarzynski_2000}. 
Following this convention, the work done by the metric $\dot{W}_t^{(g)}$ is determined via the ensemble average $\langle\cdot\rangle$ over all $(x,{\pi})$ as
\begin{equation}
    \dot{W}_t^{(g)}\coloneqq\left\langle\frac{1}{2m}\partial_t(g_t)^{\mu\nu}{\pi_\mu} {\pi_\nu} \right\rangle.
    \label{eq:diff_kinetic}
\end{equation}
This induces a change in the kinetic energy $E_t^\mathrm{kin}={\frac{1}{2m}(g_t)^{\mu\nu}\pi_\mu\pi_\nu}$. In the overdamped regime, the momentum instantaneously relaxes to the Maxwell distribution {at each $x$, for which the equipartition
theorem $\langle\pi_\mu\pi_\nu\rangle_{\mathrm M}=m k_\B T(g_t)_{\mu\nu}$ holds for the ensemble
average with respect to the Maxwell distribution $\langle\cdot\rangle_{\rm M}$.}
{Applying this relation in Eq.~(\ref{eq:diff_kinetic}) gives $\dot{W}_t^{(g)}=(k_\B T/2)\langle(g_t)_{\mu\nu}\partial_t(g_t)^{\mu\nu}\rangle$, where the remaining average is over position. The relation $(g_t)_{\mu\nu}\partial_t(g_t)^{\mu\nu}=-(g_t)^{\mu\nu}\partial_t(g_t)_{\mu\nu}$ then gives $\dot{W}_t^{(g)}=\dot{G}_t$.}
{Since} the mean kinetic energy remains constant at $\langle E_t^\mathrm{kin}\rangle =\frac{d}{2}k_\B T$, {the corresponding contribution to the heat absorbed by the system is $-\dot{W}_t^{(g)}$.}

\begingroup
Consider an underdamped model in which the time-dependent manifold is embedded in the ambient space, i.e., a higher dimensional Euclidean space.
The above dynamics and the expressions for the average heat and work can also be obtained by taking the overdamped limit of this model (see Ref.~\cite{LeeKim2026} for a two-dimensional manifold embedded in three-dimensional Euclidean space with isotropic tangential friction, and see Sec.~\ref{sec:supp_embedded} {in SM} for the general case). 
In this model, the intrinsic heat, work, and entropy production on $M$ can be separated from the contributions due to the manifold's normal motion relative to the bath.
We emphasize, however, that our framework is self-contained within the overdamped description and does not rely on this discussion.
\par
\endgroup
\parhead{{Naive definitions}}The necessity of the correction $\dot{G}_t$ becomes {clearer} by examining the case in which it is neglected. If one naively identifies heat and work solely from the time derivative of the energy~(\ref{eq:1st_law_od_pre}) as
\begin{align}
{\dot{Q}_t^\mathrm{naive}}&\coloneqq \int_M (J_t)^\mu[\nabla_\mu \phi_t]\dvolg,\\
{\dot{W}_t^\mathrm{naive}}&\coloneqq \int_M\rho_t(\partial_t\phi_t)\dvolg,
\label{eq:def_QW_naive_rate}
\end{align}
the corresponding entropy production reads
\begin{align}
  \dot{\Sigma}_t^\mathrm{naive}&{\coloneqq}\dot{S}_t^\mathrm{sys}-\frac{1}{k_\B T}{\dot{Q}_t^\mathrm{naive}}\\
  &=\int_M\left\|\nab\ln\frac{\rho_t}{\rho_t^\mathrm{eq}}\right\|_{D_t}^2\rho_t\dvolg+\int_M\rho_t\,\partial_t\dvolg.
\end{align}
That is, when the space contracts and the second term becomes negative with an absolute value exceeding the first term, the second law is violated. We thus see that the correction~(\ref{eq:G_od_rate}) is essential for preserving the 
second law.\par
\begingroup
\parhead{Example 1}
{As an experimentally feasible example, we consider} a bent colloidal chain whose end-to-end distance $L_t$ is controlled by optical traps, as considered in Ref.~\cite{Stuij2019} (Fig.~\ref{fig:chain}~(a)). We fix each bond length at \(a\). For weak bending, the internal transverse displacement vector $y=(y_1,\ldots,y_{N-2})^\top$ is confined to the ellipsoid $y^\top Ay=2\delta_t$, where $\delta_t=(N-1)a-L_t$ and $A$ is the symmetric $(N-2)\times(N-2)$ matrix defined by $y^\top Ay=\sum_{j=1}^{N-1}(y_j-y_{j-1})^2/a$, with $y_0=y_{N-1}=0$.
Let $M$ be the ellipsoid at time $0$.
Then, $g_t$ is the metric on the ellipsoid at time $t$ pulled back to $M$.
Taking the pullback along the normal direction to the ellipsoid, we have $\partial_t g_t=2\dot\delta_tA/{|Ay|^2}$, which yields
\begin{align}
  \dot G_t=-\kB T\dot\delta_t
  \left\langle\frac{\tr(P_yA)}{|Ay|^2}\right\rangle_t,
  \label{eq:chainG}
\end{align}
where $P_y$ denotes the orthogonal projection onto the tangent space of the ellipsoid at $y$, and the brackets denote expectation over $y$ at time $t$. Since $A$ is positive definite, $\dot G_t$ and $\dot\delta_t$ have opposite signs.

For the numerical calculation, we take six beads with equal isotropic friction $(\gamma_t)_{\mu\nu}=\zeta(g_t)_{\mu\nu}$. Their four internal displacements define a three-dimensional configuration manifold. We widen the chain and decrease $\delta_t$ according to
$\delta_t=\delta_0+(\delta_\tau-\delta_0)t/\tau$
for $\delta_\tau<\delta_0$. The weak-bending potential is given by $\phi_t(y)=\tfrac{\kappa}{2}|Ay|^2$ with $\kappa$ being the stiffness of the chain. We take $\rho_0\propto\exp[-\phi_0/(\kB T)]$ with respect to the induced volume measure at time $0$.

\begin{figure}[t]
  \centering
  {\includegraphics[width=0.9\columnwidth]{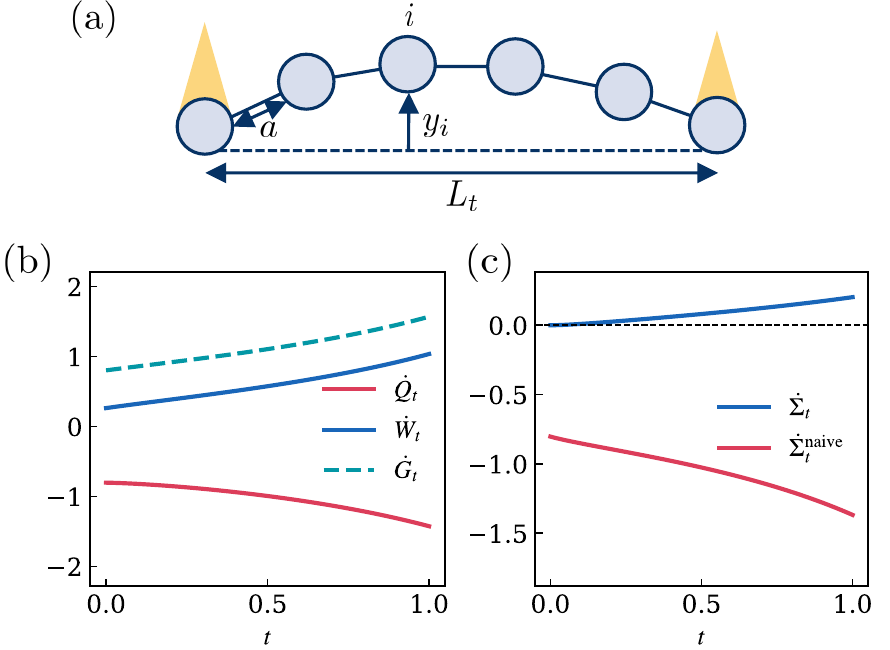}}
  \caption{(a) Schematic of the colloidal chain with its ends held in optical traps.
  (b) Time evolution of $\dot Q_t$, $\dot W_t$, and $\dot G_t$. In units $a=\kB T=\tau=1$, the parameters are $N=6$, $\zeta=60$, $\kappa=12$, $\delta_0=0.040$, and $\delta_\tau=0.022$.
  (c) Comparison of $\dot\Sigma_t$ and $\dot\Sigma_t^{\mathrm{naive}}$. The entropy production rate is nonnegative, while the naive entropy production rate is negative.}
  \label{fig:chain}
\end{figure}

We numerically solve the overdamped Fokker--Planck equation (\ref{eq:conservation}) with the current (\ref{eq:def_J_setup}) on the evolving ellipsoid. Fig.~\ref{fig:chain}~(b) shows the heat, work, and geometric correction rates. As the end-to-end distance increases, positive work is done on the chain, and the chain releases heat. Fig.~\ref{fig:chain}~(c) compares {$\dot{\Sigma}_t$ with $\dot\Sigma_t^{\mathrm{naive}}$, demonstrating that the correct entropy production rate $\dot\Sigma_t$ is consistent with the second law, whereas the naive one $\dot\Sigma_t^{\mathrm{naive}}$ yields negative values, violating the second law.}
\par\endgroup

\parhead{Example 2}In order to discuss an interesting correspondence with the entropy structure in {mathematics}, we now consider the case in which {$M$ is closed} and the metric evolves according to Ricci flow. {We take $\phi_t=0$ and specialize to $(D_t)^{\mu\nu}=(g_t)^{\mu\nu}$.} Ricci flow is a time evolution of the metric that smooths out curvature inhomogeneities, which has been extensively studied in geometric analysis~\cite{Hamilton_1982,perelman2002entropy,Topping_2006}{, and also appears in physical models such as epithelial morphogenesis~\cite{VafaMahadevan2022}.} In mathematics, it is standard to consider the {``conjugate heat equation''} that describes the time-reversal of physically feasible diffusion process~\cite{Topping_2006}. {To make contact with this mathematical setting while considering the forward heat equation, we take the metric to evolve according to backward Ricci flow}
\begin{align}
  (\partial_t g_t)_{\mu\nu} = 2(\mathrm{Ric}_t)_{\mu\nu},\label{eq:ricci_flow}
\end{align}
where $(\mathrm{Ric}_t)_{\mu\nu}$ is the Ricci curvature tensor with respect to $g_t$. In this case, the correction term $\dot{G}_t$ can be written as the average of the scalar curvature $R_t\coloneqq (g_t)^{\mu\nu}(\mathrm{Ric}_t)_{\mu\nu}$,
\begin{align}
  \dot{G}_t
  &=-k_\B T\int_M \rho_tR_t\dvolg.
  \label{eq:God_ricci}
\end{align}
Setting $h_t(x)\coloneqq-\ln\rho_t(x)$, we have
\begin{align}
    \dot{S}_t^\mathrm{sys}&=\int_M \left({\|\nabla h_t\|_{g_t}}^2+R_t\right)e^{-h_t}\dvolg,
\label{eq:Sod_ricci_perelman}
\end{align}
This expression (\ref{eq:Sod_ricci_perelman}) coincides with Perelman's $\mathcal{F}$-functional~\cite{Topping_2006} that was used for the proof of the Poincar\'{e} conjecture~\cite{perelman2002entropy}, given by
\begin{align}
    \mathcal{F}[g_t,h_t]\coloneqq\int_M\left({\|\nabla h_t\|_{g_t}}^2+R_t\right)e^{-h_t}\dvolg.\label{eq:Perelman_F}
\end{align}
Therefore, our framework clarifies a physical interpretation of Perelman's $\mathcal{F}$-functional~(\ref{eq:Perelman_F}), where the first and the second terms in the integral corresponds to the entropy production $\dot{\Sigma}_t$ and the correction term $-\dot{G}_t$, respectively.

\parhead{Discussion}In this Letter, {we have formulated stochastic thermodynamics for} an overdamped Langevin system under a time-dependent metric tensor $g_t$.
Our main contribution is incorporating arbitrary time dependence of the metric and identifying the explicit correction~(\ref{eq:G_od_rate}) to the energy balance.
This nontrivial {correction} cannot be captured by {naively} decomposing the time derivative of the energy, and can be viewed as a consequence of the definition of entropy production~{(\ref{eq:sigma_od_relent})} and thermodynamic consistency.
{While our results can be formulated as a closed theory within the overdamped regime, they are also consistent with the limit taken from the underdamped regime (Sec.~D {in} SM).}
This framework not only allows us to evaluate the thermodynamic costs for stochastic processes on time-evolving spaces, but also provides a means of controlling dissipation through the time evolution of the background space.\par
As a side remark, we discuss the correspondence with an existing relativistic framework. {Ref.~\cite{Fingerle_2007}} considered a relativistic Brownian particle in a uniformly expanding universe and identified the bath entropy contribution from cosmic expansion, whose nonrelativistic overdamped limit corresponds to {the special case of Eq.~(\ref{eq:G_od_rate}) for uniform expansion of a sphere (see Sec.~\ref{sec:supp_sphere} {in} SM)}. Our {study}, by contrast, constructs stochastic thermodynamics for an arbitrary, generally nonuniform, time-dependent metric. This nonuniformity is necessary to describe {typical} nonrelativistic phenomena on time-evolving spaces~\cite{Naji_2007,Duncan_2015,H_kansson_2018,bell2026surface,Krause2019,Bozzini2025}, and allows one to observe curvature-dependent corrections such as Eq.~(\ref{eq:Sod_ricci_perelman}) in the case of {backward} Ricci flow. Incorporating such nonuniform formulations into relativistic frameworks is an important direction for future work.\par
Further directions include extensions to systems with nonconservative forces~\cite{Hatano_2001,Speck_2005,Ding_2022}, nonuniform temperature fields~\cite{Matsuo_2000,Celani_2012,Marino_2016}, and nonlinear friction~\cite{Sarracino_2013}. {For systems in a background fluid flow~\cite{Speck_2008,Lan_2015,Wu_2024}, one could extend the flow model of Sec.~\ref{sec:supp_embedded} in SM to include the response of the fluid to the system's dynamics.} {Another direction is to develop a formulation based on probabilities of trajectories. For compact manifolds without boundary, such a formulation is available under suitable regularity conditions (see Sec.~\ref{sec:supp_frozen_kernel} in SM). This provides a starting point for formulating response relations and fluctuation theorems on time-evolving curved spaces.} {Beyond theoretical developments, experimental verification is also important.
For instance, colloidal particles or passive surfactants on deforming interfaces~\cite{XuZhao2003,Duncan_2015,Lenz_2011,memDziuk2007}, particles on contracting closed curves~\cite{GhoshMarquezLago2015}, and Brownian particles in growing media with harmonic confinement~\cite{LeVotYusteAbad2019} or reflecting boundaries~\cite{SmithYates2021} fall within the scope of our framework and could serve as platforms for such verification.}
{\parhead{Note added}After submission of our work, we became aware of Ref.~\cite{LeeKim2026}, which identifies the geometric correction terms for two-dimensional systems embedded in three-dimensional Euclidean space. While their approach is based on embedded underdamped dynamics, our derivation does not rely on embedding; it is based on thermodynamic consistency within overdamped regime and applies {to} general $d$-dimensional configuration spaces.
{
Our approach determines heat and work, including their net geometric contributions, thereby ruling out net contributions from other mechanisms of energy exchange.
This provides a thermodynamic foundation for the kinetic-energy interpretation presented around Eq.~(\ref{eq:diff_kinetic}), which Ref.~\cite{LeeKim2026} also uses in discussing an extension of their results.
}
Their paper was published on August 11, 2026, with no prior arXiv {submission}, while our results were
made publicly available on arXiv (arXiv:2606.14186) on June {15}, 2026{. We {certify}} that our work
was developed independently from theirs.
}

\parhead{Acknowledgments}
{We thank Isaac Layton and Kaito Tojo for valuable discussions.} This work is supported by JST ERATO Grant No.~JPMJER2302, Japan. R.N. is supported by the World-leading Innovative Graduate Study Program for Materials Research, Information, and Technology (MERIT-WINGS) of the University of Tokyo and by JSPS KAKENHI Grant No.~JP26KJ0820. 
{S.S. is supported by JSPS KAKENHI Grant Number JP26K17047. }
{A.T. is supported by JSPS KAKENHI Grant Number 24K21513.} T.S. is supported by JST CREST Grant No.~JPMJCR20C1 and by the Institute of AI and Beyond of the University of Tokyo.

\begingroup
\endgroup
\makeatletter
\edef\letterreferencecount{\arabic{NAT@ctr}}

\makeatother

\clearpage
\onecolumngrid

\setcounter{secnumdepth}{3}
\setcounter{section}{0}
\setcounter{subsection}{0}
\setcounter{equation}{0}
\setcounter{figure}{0}
\setcounter{table}{0}
\setcounter{page}{1}

\renewcommand{\thesection}{\Alph{section}}
\renewcommand{\thesubsection}{\Alph{section}.\arabic{subsection}}
\renewcommand{\theequation}{S\arabic{equation}}
\renewcommand{\theHequation}{supp.\arabic{equation}}
\renewcommand{\thefigure}{S\arabic{figure}}
\renewcommand{\theHfigure}{supp.\arabic{figure}}
\renewcommand{\thetable}{S\arabic{table}}
\renewcommand{\thepage}{S\arabic{page}}

\begin{center}
    \textbf{Supplemental material for ``Stochastic Thermodynamics on Time-Evolving Curved Spaces''}
\end{center}
\section{Mathematical remarks}
{In the main text}, we assume that {the topology and the differentiable structure} of the configuration space {are} preserved in time, so that the configuration spaces at different times are diffeomorphic. The physical change of the geometric structure is then encoded in the time evolution of the metric, and we may assume without loss of generality that the manifold $M$ is fixed. {Specifically}, given a family of Riemannian manifolds $(M_t, g'_t)$ at each time $t$, an equivalent representation $(M, g_t)$ is obtained by pulling them back to the manifold $M = M_0$ at $t = 0$.\par
We assume that the probability measure $\mu_t$ giving the particle's position at time $t$ is absolutely continuous with respect to $\dvol_{g_t}$. {The domain $U$ in the probability conservation 
$\frac{\dd}{\dd t}\int_U\rho_t\,\dvol_{g_t} = -\int_{\partial U}(J_t)^\mu n_\mu\,\dd A_{g_t}$ is assumed to be smooth and bounded. By applying divergence theorem $\int_{\partial U}(J_t)^\mu n_\mu\,\dd A_{g_t}=\int_U\nabla_\mu(J_t)^\mu\dvol_{g_t}$, we obtain Eq.~(\ref{eq:conservation}) in the main text.}\par
When the manifold $M$ is not closed, we impose that the probability current vanishes at the boundary or at infinity. Then, for any integrable {test function} $\varphi$, integration by parts yields
\begin{align}
  \int_M (J_t)^\mu(\nabla_\mu\varphi)\,\dvol_{g_t} 
  = -\int_M [\nabla_\mu (J_t)^\mu]\,\varphi\,\dvol_{g_t}.
\end{align}\par
{We assume that the subset $U \subset M$ in the definition of the probability measure $K_t^\epsilon(U|x)$ in {Sec.~\ref{sec:supp_frozen_kernel}} is measurable. We also note that integrals involving $\ln\rho_t$ are evaluated only on the region where $\rho_t(x)>0$.}

\section{Coordinate invariance}
\label{sec:supp_coordinate_invariance}

In this section, we verify that the density $\rho_t$ defined by Eq.~(\ref{eq:def_rho}) in 
the main text is invariant under coordinate transformations.\par
Consider two time-independent {local coordinate systems $(x^\mu)$ and $(\tilde{x}^\mu)$ around an arbitrary point of $M$}. Denoting the coordinate transformation as $x^\mu = x^\mu(\tilde{x})$, we define the Jacobian 
matrix
\begin{align}
  A^\mu{}_\nu\coloneqq
  \frac{\partial x^\mu}{\partial \tilde{x}^\nu}.
\end{align}
{The components of the metric} in the {coordinate system} $(\tilde{x}^\mu)$ can be written as
\begin{align}
  {(\tilde{g}_t)}_{\alpha\beta}{(\tilde{x})}=A^\mu{}_\alpha A^\nu{}_\beta(g_t)_{\mu\nu}(x),\label{eq:metric_transformation}
\end{align}
so that
\begin{align}
  \det{(\tilde{g}_t)}_{\alpha\beta}(\tilde{x})
  =\left[\det\left(A^\mu{}_\nu\right)\right]^2\det(g_t)_{\mu\nu}(x).
\end{align}
On the other hand, since
\begin{align}
  \dd x
  =
  \left|
    \det\left(A^\mu{}_\nu\right)
  \right|
  \dd \tilde{x},
\end{align}
the Riemannian volume measure satisfies
\begin{align}
  \dd\widetilde{\operatorname{vol}}_{g_t}
  &:=
  \sqrt{\det{(\tilde{g}_t)}_{\alpha\beta}}\,
  \dd \tilde{x}
  \\
  &=
  \sqrt{\det(g_t)_{\mu\nu}}\,
  \dd x
  \\
  &=
  \dvol_{g_t}.
\end{align}
This implies that $\dvol_{g_t}$ is independent of the choice of coordinates. Since the probability measure $\dd\mu_t$ is also coordinate-independent,
\begin{align}
  \dd\mu_t
  =
  \rho_t(x)\,\dvol_{g_t}
  =
  \tilde{\rho}_t(\tilde{x})\,\dd\widetilde{\operatorname{vol}}_{g_t}
\end{align}
holds. Consequently, we obtain
\begin{align}
  \tilde\rho_t(\tilde{x})
  =
  \rho_t(x),
\end{align}
which shows that $\rho_t$ is invariant under coordinate transformations.\par
Comparing $\rho_t$ with the density with respect to $\dd x$,
\begin{align}
  P_t(x)\coloneqq\frac{\dd\mu_t}{\dd x},
\end{align}
we have the relation
\begin{align}
  P_t(x)=\rho_t(x)\sqrt{\det(g_t)_{\mu\nu}}.
\end{align}
This expression reproduces the correction proposed in Ref.~[22].
\par
We next verify that the correction term $\dot{G}_t$ associated with the time 
variation of the metric is also coordinate-invariant. {By taking time derivative of Eq.~(\ref{eq:metric_transformation}), we obtain}
\begin{align}
  \partial_t{(\tilde{g}_t)}_{\alpha\beta}=A^\mu{}_\alpha A^\nu{}_\beta\partial_t(g_t)_{\mu\nu}.
\end{align}
Furthermore, the transformed inverse metric ${(\tilde{g}_t)}^{\alpha\beta}$, defined by 
$(g)'^{\alpha\beta}(g)'_{\beta\gamma}=\delta^\alpha{}_\gamma$, satisfies
\begin{align}
  {(\tilde{g}_t)}^{\alpha\beta}{(\tilde{x})}={(A^{-1})^\alpha{}_\mu (A^{-1})^\beta{}_\nu}(g_t)^{\mu\nu}(x),
\end{align}
where
\begin{align}
    (A^{-1})^\mu{}_\nu\coloneqq\frac{\partial\tilde{x}^\mu}{\partial x^\nu}.
\end{align}
Combining these with $A^\mu{}_\kappa(A^{-1})^\kappa{}_\nu=\delta^\mu{}_\nu$, we find
\begin{align}
  \operatorname{tr}_{g_t'}[\partial_tg'_t]&={(\tilde{g}_t)}^{\alpha\beta}\partial_t{(\tilde{g}_t)}_{\alpha\beta}\\
  &=(g_t)^{\mu\nu}\partial_t(g_t)_{\mu\nu}\\
  &=\operatorname{tr}_{g_t}[\partial_tg].
\end{align}
Hence $\operatorname{tr}_{g_t}[\partial_tg]$ is coordinate-invariant. Combined with 
the coordinate invariance of $\rho_t$ and $\dvol_{g_t}$, it follows that $\dot{G}_t$ is 
also coordinate-invariant.

\begingroup
\section{Formulation based on forward and backward processes}
\label{sec:supp_frozen_kernel}
\endgroup

{In this section, we formulate the entropy production rate defined by
Eq.~(\ref{eq:sigma_od_relent}) in terms of forward and backward processes.}

\begingroup
We first describe the probability measure on trajectories for the actual time-dependent dynamics.
On a compact manifold without boundary, if the coefficients are smooth on $M\times[0,\tau]$, the generator $L_t$ defined below determines a unique diffusion process for each initial distribution, and hence determines a probability measure on $C([0,\tau],M)$, the space of continuous trajectories~\cite{CoulibalyPasquier2011,Arnaudon_2010}.
On noncompact manifolds, non-explosion must additionally be ensured to obtain a probability measure on continuous trajectories over the entire interval; see Ref.~\cite{Kuwada_2010} for sufficient conditions in the setting considered there.

Let $P_{s,t}(\dd y|x)$ denote the transition kernel for this time-dependent
process. Given an initial probability measure $\mu_0$, for
$0=t_0<t_1<\cdots<t_n\leq\tau$, its finite-dimensional
distributions are
\[
 \mathbb P(X_{t_0}\in\dd x_0,\ldots,X_{t_n}\in\dd x_n)
 =\mu_0(\dd x_0)\prod_{i=0}^{n-1}
 P_{t_i,t_{i+1}}(\dd x_{i+1}|x_i).
\]
These consistent distributions characterize the probability measure on
continuous trajectories. When written as a density, the kernel takes the
form
\[
 P_{s,t}(\dd y|x)=p(s,x;t,y)\,\dvol_{g_t}(y),\qquad s<t.
\]
Thus, the density is taken with respect to the volume measure at the
arrival time. The kernel $P_{s,t}$ includes the variation of the
coefficients throughout $[s,t]$; it is distinct from the auxiliary kernel
$K_t^\epsilon$ used below, for which the coefficients are held fixed.
\endgroup

Fixing $g_t${,} $\phi_t${, and $D_t$}, we consider the time evolution of the probability
density from time $t$ to $t+\epsilon$ according to
Eqs.~(\ref{eq:conservation}) and (\ref{eq:def_J_setup}).
Let $K_t^\epsilon(U|x)$ denote the {conditional probability} for the particle
at $x$ at time $t$ to be in $U\subset M$ at time $t+\epsilon$.
{To define $K_t^\epsilon$ explicitly,} with the time parameter relabeled as $s$,
we consider the time evolution
\begin{align}
    \partial_s\rho_s&=\nabla_\mu\left[(D_t)^{\mu\nu}\left(\nabla_\nu\rho_s-\frac{\rho_s}{k_\B T}(f_t)_\nu\right)\right]\\
    &=\nabla_\mu\left[\rho_t^{\rm eq}(D_t)^{\mu\nu}
  \nabla_\nu
  \left(
    \frac{\rho_s}{\rho_t^{\rm eq}}
  \right)\right].
\end{align}
Since $g_t$ is fixed, the factors $\dvol_{g_t}$ on both sides of 
Eq.~(\ref{eq:conservation}) cancel. {We define the generator of this time evolution} as the operator $L_t$ satisfying, for any sufficiently smooth function $a$ on $M$,
\begin{align}
    \frac{\dd}{\dd s}\int_Ma(x)\rho_s(x)\dvol_{g_t}(x)\eqqcolon\int_ML_t[a](x)\rho_s(x)\dvol_{g_t}(x).
\end{align}
Rearranging {this}, we obtain
\begin{align}
  L_t [a]
  =
  (\rho_t^{\rm eq})^{-1}
  \nabla_\mu
  \left[
    \rho_t^{\rm eq}(D_t)^{\mu\nu}\nabla_\nu\,a
  \right].
\end{align}
The measure $K_t^\epsilon$ is then defined via the semigroup $e^{\epsilon L_t}$ 
by requiring that, for any sufficiently {regular} function $a$ on $M$,
\begin{align}
  e^{\epsilon L_t}[a](x)
  \eqqcolon
  \int_M a(y)\,K_t^\epsilon(\dd y|x).
\end{align}

We define the endpoint measure of this time evolution as 
$\dd\eta_{t,\epsilon}(y)\coloneqq\int_{x\in M}K_t^\epsilon(\dd y|x)\dd\mu_t(x)$.
For the forward and backward processes, we define the probability measures on $M\times M$ 
with the initial point $x$ and the final point $y$ as
\begin{align}
    \dd\mathbb{P}_{t,\epsilon}^\mathrm{F}(x,y)&\coloneqq\dd\mu_t(x)K_t^\epsilon(\dd y|x),\\
    \dd\mathbb{P}_{t,\epsilon}^\mathrm{B}(x,y)&\coloneqq\dd\eta_{t,\epsilon}(y)K_t^\epsilon(\dd x|y).
\end{align}
{The entropy production rate $\dot{\Sigma}_t$ can then be written as}
\begin{align}
    \dot{\Sigma}_t\mathrel{{=}}\lim_{\epsilon\searrow 0}\frac{1}{\epsilon}S\left(\mathbb{P}_{t,\epsilon}^\mathrm{F}||\mathbb{P}_{t,\epsilon}^\mathrm{B}\right).\label{eq:def_EP}
\end{align}
{This formulation} is a natural extension of the definition in Ref.~\cite{CaiWangZhao2024FluctuationTheoremRM} to the case of a time-dependent metric.\par
We next discuss detailed balance, {which establishes the equivalence of this
formulation to the definition~(\ref{eq:sigma_od_relent}) in the main text.}
The instantaneous equilibrium measure $\dd{\mu_t^{\rm eq}}(x)=\rho_t^{\mathrm{eq}}(x)\dvol_{g_t}$ 
and the transition kernel $K_t^\epsilon(\dd y|x)$ defined above satisfy the detailed 
balance condition
\begin{align}
    \dd{\mu_t^{\rm eq}}(x)K_t^\epsilon(\dd y|x)=\dd{\mu_t^{\rm eq}}(y)K_t^\epsilon(\dd x|y),\label{eq:detailed_balance}
\end{align}
which corresponds to the vanishing of the probability current at the instantaneous equilibrium distribution.
{Substituting this into Eq.~(\ref{eq:def_EP}), we recover
Eq.~(\ref{eq:sigma_od_relent}). A proof of Eq.~(\ref{eq:detailed_balance})
and of this equivalence is given below.}

{We now} show that the transition kernel satisfies detailed balance with respect to the instantaneous equilibrium measure, and that {the forward/backward formulation is equivalent to the main-text definition of entropy production}.\par
We first show that for any sufficiently {regular} functions $a,b$,
\begin{align}
  \int_M a\,L_t[b]\,\dd{\mu_t^{\rm eq}}
  &=
  \int_M a\, \nabla_\mu
  \left(
    \rho_t^{\rm eq}(D_t)^{\mu\nu}\nabla_\nu b
  \right)
  \dvolg\\
  &=-\int_M \rho_t^{\rm eq} (D_t)^{\mu\nu} (\nabla_{x^\mu}a)(\nabla_{x^\nu}b) \dvolg\\
  &=-\int_M \rho_t^{\rm eq}(D_t)^{\mu\nu}(\nabla_{x^\mu}b)(\nabla_{x^\nu}a)\dvolg\\
  &=\int_M b\,L_t [a]\,\dd{\mu_t^{\rm eq}}.
\end{align}
Consequently, the semigroup $e^{\epsilon L_t}$ also satisfies
\begin{align}
  \int_M a\,e^{\epsilon L_t}[b]\,\dd{\mu_t^{\rm eq}}
  =
  \int_M b\,e^{\epsilon L_t}[a]\,\dd{\mu_t^{\rm eq}}.
\end{align}
Setting $a=\mathds{1}_A$ and $b=\mathds{1}_B$ for arbitrary {measurable} $A,B\subset M$, we obtain
\begin{align}
  \int_A \dd{\mu_t^{\rm eq}}(x)\,K_t^\epsilon(B|x)
  =
  \int_B \dd{\mu_t^{\rm eq}}(y)\,K_t^\epsilon(A|y),
\end{align}
which is equivalent to the equality of measures on $M\times M$,
\begin{align}
  \dd{\mu_t^{\rm eq}}(x)K_t^\epsilon(\dd y|x)
  =
  \dd{\mu_t^{\rm eq}}(y)K_t^\epsilon(\dd x|y).
\end{align}
This establishes the detailed balance condition.\par
We next show that {the formulation above} via the forward process 
$\dd\mathbb P^{\rm F}_{t,\epsilon}(x,y)=\dd\mu_t(x)K_t^\epsilon(\dd y|x)$ and the 
backward process $\dd\mathbb P^{\rm B}_{t,\epsilon}(x,y)=\dd\eta_{t,\epsilon}(y)K_t^\epsilon(\dd x|y)$ 
{agrees with the definition~(\ref{eq:sigma_od_relent}) in the main text}.
From the detailed balance condition, we have
\begin{align}
  \frac{
    \dd\mathbb P^{\rm F}_{t,\epsilon}(x,y)
  }{
    \dd\mathbb P^{\rm B}_{t,\epsilon}(x,y)
  }
  =
  \frac{
    \dd\mu_t/\dd{\mu_t^{\rm eq}}(x)
  }{
    \dd\eta_{t,\epsilon}/\dd{\mu_t^{\rm eq}}(y)
  }.
\end{align}
Therefore,
\begin{align}
  S\left(
    \mathbb P^{\rm F}_{t,\epsilon}
    \middle\|
    \mathbb P^{\rm B}_{t,\epsilon}
  \right)
  &=
  \int_{M\times M}
  \ln
  \frac{
    \dd\mathbb P^{\rm F}_{t,\epsilon}
  }{
    \dd\mathbb P^{\rm B}_{t,\epsilon}
  }
  \dd\mathbb P^{\rm F}_{t,\epsilon}\\
  &=
  \int_{M\times M}
  \ln\frac{\dd\mu_t(x)}{\dd{\mu_t^{\rm eq}}(x)}\,
  \dd\mathbb P^{\rm F}_{t,\epsilon}(x,y)
  -
  \int_{M\times M}
  \ln\frac{\dd\eta_{t,\epsilon}(y)}{\dd{\mu_t^{\rm eq}}(y)}\,
  \dd\mathbb P^{\rm F}_{t,\epsilon}(x,y).
\end{align}
Since the $x$-marginal of the forward two-time measure is $\mu_t$ and the $y$-marginal 
is $\eta_{t,\epsilon}$, we obtain
\begin{align}
  S\left(
    \mathbb P^{\rm F}_{t,\epsilon}
    \middle\|
    \mathbb P^{\rm B}_{t,\epsilon}
  \right)
  =
  S(\mu_t\|{\mu_t^{\rm eq}})
  -
  S(\eta_{t,\epsilon}\|{\mu_t^{\rm eq}}).
\end{align}
We now take the $\epsilon\to0$ limit.
For any {sufficiently regular} function $a$,
\begin{align}
  \int_M a\,\dd\eta_{t,\epsilon}
  &=
  \int_M e^{\epsilon L_t} [a]\,\dd\mu_t
  \\
  &=
  \int_M a\,\dd\mu_t
  +
  \epsilon\int_M (L_t a)\,\dd\mu_t
  +
  o(\epsilon).
\end{align}
On the other hand, for the solution $\mu_s$ of the Fokker--Planck 
equation~(\ref{eq:conservation}),
\begin{align}
  \left.
  \frac{\dd}{\dd s}
  \int_M a\,\dd\mu_s
  \right|_{s=t}
  =
  \int_M (L_t a)\,\dd\mu_t.
\end{align}
Therefore,
\begin{align}
  \eta_{t,\epsilon}
  =
  \mu_t
  +
  \epsilon
  \left.
  \frac{\dd\mu_s}{\dd s}
  \right|_{s=t}
  +
  o(\epsilon).
\end{align}
Hence,
\begin{align}
  \dot\Sigma_t=
  \lim_{\epsilon\searrow0}
  \frac{1}{\epsilon}
  S\left(
    \mathbb P^{\rm F}_{t,\epsilon}
    \middle\|
    \mathbb P^{\rm B}_{t,\epsilon}
  \right)
  =
  -
  \left.
  \frac{\dd}{\dd s}
  S(\mu_s\|{\mu_t^{\rm eq}})
  \right|_{s=t},
\end{align}
which is the relative entropy representation used in the main text.\par
\addvspace{\baselineskip}

\begingroup
\let\dd\differential
\let\standardexp\exp
\def\exp{\opbraces{\standardexp}}
\makeatletter
\c@secnumdepth=2\relax
\def\p@subsection{}
\makeatother
\section{Realization of overdamped thermodynamics on a time-evolving manifold \\ from an embedded underdamped model}
\label{sec:supp_embedded}

In this section, we show how the intrinsic dynamics and thermodynamics on a time-evolving manifold $(M,g_t)$ formulated in the main text can be obtained from the overdamped limit of an underdamped model of constrained motion in Euclidean space.
After obtaining the equations of motion and thermodynamics of the ambient constrained system in Secs.~\ref{sec:appD-thermal-contact}--\ref{sec:appD-ambient-thermodynamics}, we construct its canonical description on a fixed manifold $M$ in Sec.~\ref{sec:appD-intrinsic-thermodynamics}.
We then formulate thermodynamics using the dynamics on $M$ alone and compare it with that of the ambient constrained system.
This comparison separates the ambient heat, work, and entropy production into the intrinsic contributions and those associated with the normal motion of the manifold relative to the bath.
In the overdamped limit, the intrinsic contributions to heat and work
reproduce Eqs.~(\ref{eq:Q_od_rate}) and~(\ref{eq:W_od_rate})
of the main text, respectively.

We note, however, that the formulation on a time-evolving manifold does not rely on an embedding model and also applies to abstract configuration spaces, including flat tori, even when no natural realization as constrained motion in physical space is available.

\subsection{
  Underdamped Langevin equation in the ambient space
}
\label{sec:appD-thermal-contact}

In the Euclidean ambient space $\mathbb{R}^N$ with flat metric $\delta_{\alpha\beta}$, we denote the particle's position by capital $\bm{X} \in \mathbb{R}^N$.
For arbitrary vectors $V^\alpha, W^\beta$, we write its inner product as $\bm{V}^T\bm{W} \coloneqq \delta_{\alpha\beta} V^\alpha W^\beta$.
The Lagrangian of a particle in the Euclidean space is given by
\begin{equation}
  \mathcal{L}_{\mathrm{amb}} \coloneqq \frac{m}{2} \dot{\bm X
  }^T \dot{\bm X} - U_t(\bm X).
\end{equation}
The impulse delivered by the bath during $\dd{t}$ has a deterministic part and a random part, which we denote by $\bm{F}_{\mathrm{amb,bath}} \dd{t}$ and $\sum_{r} \bm{\sigma}_{\mathrm{amb},r} \circ \dd{B_t^{(r)}}$, respectively.
Here, $B_t^{(r)} \ (r=1,2,...)$ denote independent Wiener processes, and $\circ$ denotes the Stratonovich product~\cite[Sec.~3.5]{RussoVallois2007}.
Then, the equation of motion for this system is obtained via the Euler--Lagrange equation with nonconservative forces:
\begin{equation}
  \left( \dv{}{t} \pdv{\mathcal{L}_{\mathrm{amb}}}{\dot{\bm X}} - \pdv{\mathcal{L}_{\mathrm{amb}}}{\bm X} \right) \dd{t} = \left( \bm{F}_{\mathrm{amb,nc}} + \bm{F}_{\mathrm{amb,bath}} \right) \dd{t} +\sum_{r} \bm{\sigma}_{\mathrm{amb},r} \circ \dd{B_t^{(r)}},
\end{equation}
where $\bm{F}_{\mathrm{amb,nc}}$ is the nonconservative force.
This gives
\begin{equation}
  m \dd{\dot{\bm X}} = \left( -\pdv{U_t}{\bm{X}} + \bm{F}_{\mathrm{amb,nc}} + \bm{F}_{\mathrm{amb,bath}} \right) \dd{t} + \sum_{r} \bm{\sigma}_{\mathrm{amb},r} \circ \dd{B_t^{(r)}}.
\end{equation}
Here, the noise amplitudes $\bm{\sigma}_{\mathrm{amb},r}$ define the momentum diffusion tensor $C_{\mathrm{amb}}^{\alpha\beta} \coloneqq \frac{1}{2} \sum_r \sigma_{\mathrm{amb},r}^\alpha \sigma_{\mathrm{amb},r}^\beta$, which is positive definite.

We consider a bath at constant temperature $T>0$ with velocity $\bm W_t(\bm X)$, and assume that $\bm F_{\mathrm{amb,nc}}$ and the noise amplitudes depend only on position and time.
To determine the relation between the deterministic and random parts of the bath force, we require the momentum dynamics due to the bath to satisfy detailed balance with respect to the Maxwell distribution~\cite{Maes_2021}.
Since the bath moves with velocity $\bm W_t(\bm X)$, this distribution involves the particle's relative velocity $\dot{\bm X}-\bm W_t(\bm X)$~\cite{Lan_2015}.
With the canonical momentum $\bm P\coloneqq m\dot{\bm X}$, it is proportional to $\exp[-\lvert\bm P-m\bm W_t(\bm X)\rvert^2/(2mk_BT)]\,\dd[N]{P}$.
The momentum current driven by the deterministic bath force must then cancel the diffusive current, which gives
\begin{equation}
  \bm F_{\mathrm{amb,bath}} = - K_{\mathrm{amb}}(\dot{\bm X}-\bm W_t(\bm X)).
\end{equation}
Here the friction tensor $K_{\mathrm{amb}}^{\alpha\beta}$ satisfies the Einstein relation $K_{\mathrm{amb}}^{\alpha\beta} = C_{\mathrm{amb}}^{\alpha\beta}/(k_BT)$~\cite{Kramers1940}.

The heat absorbed by the particle during $\dd{t}$ is the mechanical work done on it by the bath~\cite{Sekimoto_1998, Matsuo_2000}.
The velocity entering this work is the particle's velocity relative to the bath~\cite{Speck_2008}.
Thus, we obtain
\begin{equation}
  \delta{Q}_{\mathrm{amb}} \coloneqq (\dot{\bm X} - \bm{W}_t(\bm{X}))^T \circ \left( - K_{\mathrm{amb}} (\dot{\bm X} - \bm{W}_t(\bm{X})) \dd{t} + \sum_r \bm{\sigma}_{\mathrm{amb},r} \circ \dd{B_t^{(r)}} \right).
  \label{eq:appD-Qamb}
\end{equation}
Here, for simplicity, we write $\bm{Y}^T \circ \bm{Z} \coloneqq \delta_{\alpha\beta} Y^{\alpha} \circ Z^\beta$.
The work is then determined by the first law of thermodynamics: $\delta W_{\mathrm{amb}} \coloneqq \dd{E_{\mathrm{amb}}}-\delta Q_{\mathrm{amb}}$, where the energy is given by
\begin{equation}
      E_{\mathrm{amb}} \coloneqq \frac{m}{2} \dot{\bm X}_t^T \dot{\bm X}_t + U_t(\bm X_t).
\end{equation}

\subsection{Constrained dynamics on an embedded manifold}
We now constrain the particle to lie on a $d$-dimensional embedded submanifold $N_t\subset\mathbb R^N$ that can move and deform smoothly with time.
To describe the particle's motion relative to the moving constraint, we use a fixed manifold $M$ diffeomorphic to each $N_t$, for example $M=N_0$.
The smooth evolution of $N_t$ then allows us to choose a smooth family of diffeomorphisms $\bm{R}_t\colon M\to N_t$.
Each choice defines a correspondence between points at different times, $\bm{X}_t \in N_t$ and $\bm{X}_{t'} \in N_{t'}$, via $\bm{X}_{t'} = (\bm{R}_{t'}\circ\bm{R}_t^{-1})(\bm X_t)$.
We note that all such families represent the same physical constraint, since each has the same set of allowed positions $N_t = \bm{R}_t(M)$ at every time.
Thus, the choice of $\bm{R}_t$ affects only how we represent the constrained dynamics on $N_t$ as dynamics on fixed $M$.

At each fixed $t$, the metric induced on $M$ by the embedding map $\bm{R}_t$ is determined by the ambient line element $\dd{s}^2 = \dd{\bm{R}_t}^T \dd{\bm{R}_t}$.
Substituting $\dd{\bm{R}_t} = (\partial_\mu\bm{R}_t)\dd{x^\mu}$ gives the induced metric $(g_t)_{\mu\nu}\coloneqq (\partial_\mu\bm{R}_t)^T (\partial_\nu\bm{R}_t)$.
This gives a time-evolving Riemannian manifold $(M,g_t)$.

For each choice of $\bm{R}_t$, every ambient trajectory satisfying $\bm X_t\in N_t$ is represented on $M$ by $x_t\coloneqq\bm{R}_t^{-1}(\bm X_t)$.
We can therefore impose the constraint $\bm{X} \in N_t$ in the form $\bm{X} = \bm{R}_t(x)$ without loss of generality.
Depending on how the constraint is realized, the constraint mechanism may contribute to $U_t,\bm F_{\mathrm{amb,nc}}$ and, at the same $T,\bm W_t$, to the thermal friction and fluctuations $K_{\mathrm{amb}},\bm\sigma_{\mathrm{amb},r}$.
We take any such contributions to be included in the corresponding quantities.
The remaining constraint force needed to maintain $\bm X_t\in N_t$ is described by an unknown stochastic impulse $\bm\lambda(t)\dd t$ to the particle.
Its effect is incorporated by the Lagrangian with multipliers:
\begin{equation}
  \mathcal{L}
  \coloneqq \mathcal{L}_{\mathrm{amb}} + \bm{\lambda}(t)^T (\bm{X} - \bm{R}_t(x))
  = \frac{m}{2} \dot{\bm X}^T \dot{\bm X} - U_t(\bm X) + \bm{\lambda}(t)^T (\bm{X} - \bm{R}_t(x)).
\end{equation}
The Euler--Lagrange equations for $\bm{X} \in \mathbb{R}^N$ and $x \in M$ are
\begin{align}
  m \dd{\dot{\bm X}_t}
  &=
  \left[ -\pdv{U_t}{\bm{X}} + \bm{F}_{\mathrm{amb,nc}} \right] \dd{t} + \bm{\lambda}(t) \dd{t}
  - K_{\mathrm{amb}} (\dot{\bm X}_t-\bm W_t(\bm X_t)) \dd{t}
  + \sum_r \bm{\sigma}_{\mathrm{amb},r}\circ\dd{B_t^{(r)}},
  \nonumber
  \\
  0 &= \bm{v}^T(\bm\lambda(t)\dd{t}) \quad\text{for every $\bm v$ tangent to $N_t$ at $\bm X_t$}.
  \label{eq:appD-constrainedEQM}
\end{align}
The second equation follows from $0 = (\pdv*{\mathcal{L}}{x^\mu}) \dd{t} = - (\partial_\mu \bm{R}_t)^T \bm{\lambda}(t) \dd{t}$ because the vectors $\partial_\mu \bm{R}_t$ span the tangent space to $N_t$ at $\bm X_t$.
These equations of motion and the constraint $\bm{X}_t\in N_t$ are all independent of the choice of $\bm{R}_t$.
Thus, with the same ambient initial conditions and the same realization of the bath noise, the trajectory $\bm X_t$ is also independent of the choice of $\bm{R}_t$.
The heat defined in Eq.~\eqref{eq:appD-Qamb} is therefore unchanged by the choice of $\bm{R}_t$, since it is evaluated along the same ambient trajectory with the same bath impulse.

\subsection{Thermodynamics of the ambient constrained dynamics}
\label{sec:appD-ambient-thermodynamics}
We then consider the thermodynamics of the constrained dynamics~\eqref{eq:appD-constrainedEQM}.
To separate the heat associated with the motion along the embedded manifold $N_t$, we decompose the particle's velocity relative to the bath as
\begin{equation}
  \dot{\bm X_t} - \bm{W}_t(\bm X_t) = \bm{V}_t + \bm{b}_t,
\end{equation}
where $\bm V_t$ is tangential to $N_t$, and $\bm b_t$ is normal to $N_t$ in the inner product determined by the diffusion tensor $C_{\mathrm{amb}}$.
In fact, $\bm b_t$ is the value of a smooth field $\bm b_t(\bm X)$ on $N_t$, so that $\bm b_t = \bm b_t(\bm X_t)$.
To see this, at each $\bm X\in N_t$, consider velocities $\bm v$ compatible with the evolving constraint so that $\bm{X} + \bm{v} \dd{t}$ lies in $N_{t+\dd{t}}$ (to first order in $\dd{t}$).
Such velocities differ only by a vector tangent to $N_t$, because their endpoints lie on the same $N_{t+\dd t}$, and the line joining these endpoints approaches a tangent direction to $N_t$ as $\dd{t} \to 0$.
The corresponding velocities relative to the bath therefore share the same normal component in the inner product determined by $C_{\mathrm{amb}}$.
Thus, the field $\bm b_t(\bm X)$ is well-defined on $N_t$~\cite{delHoyo2016}.

By the Einstein relation, the friction force $- K_{\mathrm{amb}} \bm{b}_t$ therefore has zero tangential component.
Substituting this decomposition into Eq.~\eqref{eq:appD-Qamb} gives
\begin{align}
  \delta Q_{\mathrm{amb}}
  = \underbrace{\left[ -\bm{V}_t^T K_\mathrm{amb} \bm{V}_t \dd{t} + \sum_r (\bm{V}_t^T \bm{\sigma}_{\mathrm{amb},r}) \circ \dd{B_t^{(r)}} \right]}_{\eqqcolon \delta Q_{M}}
  + \underbrace{\left[ -\bm{b}_t^T K_\mathrm{amb} \bm{b}_t \dd{t} + \sum_r (\bm{b}_t^T \bm{\sigma}_{\mathrm{amb},r}) \circ \dd{B_t^{(r)}} \right]}_{\eqqcolon \delta Q_{b}}.
\end{align}
A benefit of this decomposition is that it separates the heat attributable to the part of the particle's relative motion that can be represented by a displacement within $N_t$ at a fixed time from the heat that must be attributed to the motion of $N_t$ relative to the bath in the ambient space.
The trajectory determines the tangential bath impulse through the momentum balance along $N_t$ and hence determines $\delta Q_M$.
To infer $\delta Q_b$ from the same trajectory, we would also need the random bath impulse along $\bm b_t$. For every $\bm v$ tangent to $N_t$ at $\bm X_t$, the cross-covariance between the random contribution to $\delta Q_b$ and the random bath impulse in the direction $\bm v$ is
\begin{equation}
  \sum_r
  (\bm{v}^T\bm\sigma_{\mathrm{amb},r})
  (\bm{b}_t^T\bm\sigma_{\mathrm{amb},r})\,\dd{t}
  = 2\bm{v}^T C_{\mathrm{amb}}\bm b_t\,\dd{t}
  = 0.
  \label{eq:appD-independence}
\end{equation}
Since the bath noise is Gaussian, the noise in $\delta Q_b$ can be represented by a Wiener process independent of those driving the motion within $N_t$.

We now quantify the irreversibility of the total system (particle and bath) through a comparison of forward and time-reversed probabilities~\cite{MaesNetocny2003}.
To retain both heat contributions, we compare the joint probabilities of the particle trajectory and its heat with those of their time reverse~\cite{Jarzynski_2000}.

For the comparison on an interval $[t,t+\Delta t]$, we consider the time reverse of the trajectory $\{\bm X_{t+s}\}_{s\in[0,\Delta t]}$ defined by
\begin{equation}
 (\Theta_*\bm X)_s\coloneqq\bm X_{t+\Delta t-s},
 \quad
 m\dv{(\Theta_*\bm X)_s}{s}=-\bm P_{t+\Delta t-s},
 \quad 0\le s\le\Delta t.
 \label{eq:appD-pathReversal}
\end{equation}
We compare the probability of obtaining the trajectory $\{\bm X_{t+s}\}_{s\in[0,\Delta t]}$ and the heat $\Delta Q_{\mathrm{amb}} \coloneqq \int_t^{t+\Delta t} \delta Q_{\mathrm{amb}}$ in the dynamics~\eqref{eq:appD-constrainedEQM} with the probability of obtaining the reversed trajectory $\{(\Theta_*\bm X)_s\}_{s=0}^{\Delta t}$ and $-\Delta Q_{\mathrm{amb}}$ in the reversed dynamics.
The equations of motion of the reversed dynamics are given by
\begin{align}
 m\dd{\dot{\tilde{\bm X}}_s}
 &=
 \left[-\pdv{\tilde U_s}{\bm X}
       +\tilde{\bm F}_{\mathrm{amb,nc}}\right]\dd s
 +\tilde{\bm\lambda}(s)\dd s
 -\tilde K_{\mathrm{amb}}
       (\dot{\tilde{\bm X}}_s-\tilde{\bm W}_s)\dd s
 +\sum_r\tilde{\bm\sigma}_{\mathrm{amb},r}
          \circ\dd\tilde B_s^{(r)},
 \nonumber\\
 &\tilde{\bm X}_s\in\tilde N_s,\quad
 0=\bm{v}^T (\tilde{\bm\lambda}(s)\dd s)
 \quad\text{for every $\bm v$ tangent to $\tilde N_s$
 at $\tilde{\bm X}_s$},
 \label{eq:appD-reverseConstrainedEQM}
\end{align}
where the constraint and coefficients are time-reversed as
\begin{equation}
 \left.
 \bigl(\tilde N,\tilde U,\tilde{\bm F}_{\mathrm{amb,nc}},
       \tilde K_{\mathrm{amb}},\tilde{\bm\sigma}_{\mathrm{amb},r},
       \tilde{\bm W}\bigr)\right|_s
 \coloneqq
 \left.
 \bigl(N,U,\bm F_{\mathrm{amb,nc}},
       K_{\mathrm{amb}},\bm\sigma_{\mathrm{amb},r},
       -\bm W\bigr)\right|_{t+\Delta t-s}.
 \label{eq:appD-reverseCoefficients}
\end{equation}
The reverse Wiener processes $\tilde B_s^{(r)}$ are mutually independent and independent of the forward Wiener processes $B_{t}^{(r)}$ and of the reverse initial state.

Once the distribution of the initial conditions for the forward process is prescribed, that for the reversed process is set to be the final distribution of the forward process with velocities reversed.
We denote by $\mathbb P^{\mathrm F}$ and $\mathbb P^{\mathrm R}$
the joint probabilities of the particle trajectory and heat generated
by the forward~\eqref{eq:appD-constrainedEQM} and reverse~\eqref{eq:appD-reverseConstrainedEQM} dynamics, respectively, with these initial distributions.
The total entropy production rate, in units of $k_B$, is defined by
\begin{equation}
    \dot{\Sigma}_{\mathrm{amb}}(t)
    \coloneqq \lim_{\Delta t \to 0} \frac{1}{\Delta t} \mathbb{E}^F\left[ \log \frac{ \mathbb{P}^{\mathrm F}\bigl[ \{\bm{X}_{t+s}\}_{s=0}^{\Delta t}, \Delta Q_{\mathrm{amb}}\bigr] }{ \mathbb{P}^{\mathrm R}\bigl[ \{(\Theta_*\bm{X})_{s}\}_{s=0}^{\Delta t}, -\Delta Q_{\mathrm{amb}}\bigr] } \right],
    \label{eq:appD-ambEPRDefinition}
\end{equation}
where $\mathbb{E}^{\mathrm F}$ denotes the average with respect to $\mathbb P^{\mathrm F}$.
Defined through the logarithmic ratio of forward and reverse joint probabilities, the stochastic total entropy production (the quantity averaged above) satisfies the detailed fluctuation theorem~\cite[Sec.~13]{ChetriteGawedzki2008}.
Its mean is the relative entropy between the forward and time-reversed probability distributions, so $\dot{\Sigma}_{\mathrm{amb}}(t)\ge0$.

We now separate the stochastic total entropy production into the part determined by the particle trajectory alone and the part that cannot be determined from it.
For an infinitesimal motion on $[t+s,t+s+\dd s]$, the prescribed trajectory determines its velocity $\dot{\bm X}$ and hence the momentum increment $\dd{\bm P} = m \dd{\dot{\bm X}}$.
Taking the inner product of Eq.~\eqref{eq:appD-constrainedEQM} with $\bm V$ then gives
\begin{equation}
 \delta Q_M
 =\bm{V}^T \circ \left[ \dd{\bm P} + \left(\pdv{U}{\bm X}-\bm F_{\mathrm{amb,nc}}\right)\dd{s} \right].
 \label{eq:appD-heatFromTrajectory}
\end{equation}
By the independence~\eqref{eq:appD-independence} established above, conditional on this motion, $\delta Q_b$ is Gaussian with mean $-\bm b^TK_{\mathrm{amb}}\bm b\,\dd s$ and variance $2\bm b^TC_{\mathrm{amb}}\bm b\,\dd s$.
Thus, for $\bm b\ne0$,
\begin{equation}
    \mathbb P^{\mathrm F}\left[\bm X_{t+s+\dd s},\delta Q_{\mathrm{amb}} \middle| \bm X_{t+s}\right]
    =
    \frac{\mathbb P^{\mathrm F}[\bm X_{t+s+\dd s}\mid\bm X_{t+s}]}{\sqrt{4\pi\bm b^TC_{\mathrm{amb}}\bm b\,\dd s}}
    \exp\!\left[
    -\frac{(\delta Q_{\mathrm{amb}}-\delta Q_M+\bm b^TK_{\mathrm{amb}}\bm b\,\dd s)^2}{4\bm b^TC_{\mathrm{amb}}\bm b\,\dd s}
    \right],
    \label{eq:appD-jointLocalProbability}
\end{equation}
where $\delta Q_M$ is evaluated from the trajectory $\bm{X}_t$ using Eq.~\eqref{eq:appD-heatFromTrajectory}.
When $\bm b=0$, the heat is fixed by the trajectory as $\delta Q_{\mathrm{amb}}=\delta Q_M$.

The corresponding reverse interval is $[\Delta t-s-\dd s,\Delta t-s]$.
Under the time-reversal in Eqs.~\eqref{eq:appD-pathReversal} and \eqref{eq:appD-reverseCoefficients}, the increment $\dd{\bm P}$ retains its sign, whereas the signs of $\bm V$ and $\bm b$ are reversed.
Thus, the part of the heat~\eqref{eq:appD-heatFromTrajectory} determined by the constrained trajectory changes sign.
Therefore, we obtain
\begin{equation}
    \log \frac{\mathbb P^{\mathrm F}\left[\bm X_{t+s+\dd s},\delta Q_{\mathrm{amb}} \middle| \bm X_{t+s}\right]}{\mathbb P^{\mathrm R}\left[(\Theta_* \bm{X})_{\Delta t - s},-\delta Q_{\mathrm{amb}} \middle| (\Theta_* \bm{X})_{\Delta t - s - \dd{s}}\right]}
    = \log \frac{\mathbb{P}^{\mathrm F}[\bm X_{t+s+\dd s}\mid\bm X_{t+s}]}{\mathbb{P}^{\mathrm R}[(\Theta_*\bm{X})_{\Delta{t} - s}\mid (\Theta_*\bm{X})_{\Delta{t} - s -\dd{s}}]} - \frac{\delta Q_b}{k_B T}.
\end{equation}
Including the initial-state probabilities in this ratio and using the average of $\delta Q_b$ conditioned on the particle's trajectory, we obtain from Eq.~\eqref{eq:appD-ambEPRDefinition}
\begin{equation}
 \dot{\Sigma}_{\mathrm{amb}}(t)
 =
 \lim_{\Delta t\to 0}\frac{1}{\Delta t}
 \mathbb{E}^{\mathrm F}\!\left[
  \log\frac{
   \mathbb{P}^{\mathrm F}\bigl[\{\bm{X}_{t+s}\}_{s=0}^{\Delta t}\bigr]
  }{
   \mathbb{P}^{\mathrm R}\bigl[\{ (\Theta_*\bm{X})_{s}\}_{s=0}^{\Delta t}\bigr]
  }
 \right]
 +
 \frac{
  \mathbb{E}^{\mathrm F}
  [\bm b_t^TK_{\mathrm{amb}}\bm b_t]
 }{k_BT}.
 \label{eq:appD-ambient-entropy-split}
\end{equation}
The first term itself has the form of a total entropy production rate, defined through the logarithmic ratio of forward and reverse probabilities of the particle trajectory alone, and is therefore nonnegative.
The second term is also nonnegative and accounts for the additional
dissipation due to the motion of $N_t$ relative to the bath.

\subsection{Dynamics and thermodynamics on $M$}
\label{sec:appD-intrinsic-thermodynamics}
We have thus formulated the equations of motion, heat, and entropy production independently of the choice of the embedding map $\bm{R}_t$.
To connect the description of the constrained dynamics and thermodynamics in the ambient space to the formulation in the main text, we now express the dynamics on the fixed manifold $M$.
We choose the correspondence between points at different times so that the particle's velocity on $M$ represents the tangential component $\bm V_t$ of its velocity relative to the bath.

For an arbitrary choice of $\bm{R}_t$, the relation
$\bm X_t=\bm{R}_t(x_t)$ gives
\begin{equation}
  \dot{\bm X}_t-\bm W_t(\bm X_t) = \dot{x}_t^\mu\partial_\mu\bm{R}_t(x_t) + \partial_t\bm{R}_t(x_t) - \bm W_t(\bm{R}_t(x_t)).
\end{equation}
This motivates us to choose $\bm{R}_t$ to satisfy
\begin{equation}
  \partial_t \bm{R}_t(x) = \bm{W}_t(\bm{R}_t(x)) + \bm{b}_t(\bm{R}_t(x)),\quad \forall x \in M.
  \label{eq:appD-embedding}
\end{equation}
With this choice, the motion on $M$ describes the particle's motion within the constraint $N_t$ relative to the bath: $\bm V_t = \dot x_t^\mu\partial_\mu\bm{R}_t(x_t)$.
Therefore, the time dependence of the induced metric $(g_t)_{\mu\nu} = (\partial_\mu\bm{R}_t)^T (\partial_\nu\bm{R}_t)$ can reflect both deformation of $N_t$ and spatial variation of the bath flow. 
In particular, the bath flow can make the induced metric time dependent even when $N_t$ is stationary.

By the definition of $\bm b_t(\bm X)$, the velocity on the right-hand side of Eq.~\eqref{eq:appD-embedding} is compatible with the evolving constraint $\bm{R}_t(x) \in N_t$.
On any time interval where the flow generated by $\bm W_t+\bm b_t$ exists for all points in both time directions, it defines diffeomorphisms $\bm{R}_t\colon M\to N_t$~\cite{delHoyo2016}.
For compact $M$ without boundary, smoothness ensures the existence of a solution $\bm{R}_t$ throughout any finite interval under consideration.

To obtain the equation of motion on $M$, we take the inner product of the ambient equation of motion~\eqref{eq:appD-constrainedEQM} with $\partial_\mu\bm{R}_t$.
The resulting equation can be written using the same Lagrangian restricted to $\bm X=\bm{R}_t(x)$:
\begin{align}
 \dd\left(\pdv{\mathcal L|_{\bm X=\bm{R}_t(x)}}{\dot x^\mu}\right)
 -\pdv{\mathcal L|_{\bm X=\bm{R}_t(x)}}{x^\mu}\dd t
 &=(\partial_\mu\bm{R}_t)^T\circ
          [m\dd{\dot{\bm X}_t}+\partial_{\bm X}U_t\dd t]
 \nonumber\\
 &=\bigl[f_\mu^{\mathrm{nc}}-(\gamma_t)_{\mu\nu}\dot x_t^\nu\bigr]\dd t
       +\sum_r[(g_t)_{\mu\nu}\sigma_r^\nu]\circ\dd B_t^{(r)},
 \label{eq:appD-EQMonM}
\end{align}
where $\phi_t\coloneqq U_t(\bm{R}_t)$ is the potential induced on $M$, and $f_\mu^{\mathrm{nc}}\coloneqq (\partial_\mu\bm{R}_t)^T \bm{F}_{\mathrm{amb,nc}}$ is the nonconservative force on $M$.
Equation~\eqref{eq:appD-EQMonM} identifies the friction tensor and noise amplitudes on $M$ as
\begin{equation}
    (\gamma_t)_{\mu\nu}
    \coloneqq 
    (\partial_\mu\bm{R}_t)^T K_{\mathrm{amb}}(\partial_\nu\bm{R}_t),
    \quad
    \sigma_r^\mu 
    \coloneqq (g_t)^{\mu\nu} (\partial_\nu\bm{R}_t)^T\bm{\sigma}_{\mathrm{amb},r}.
    \label{eq:appD-induced-K}
\end{equation}
Here $(g_t)^{\mu\nu}$ is the inverse induced metric.
The corresponding momentum diffusion tensor is
\begin{equation}
    (C_t)_{\mu\nu}
    \coloneqq 
    \frac{1}{2} \sum_r (g_t)_{\mu\kappa}\sigma_r^\kappa (g_t)_{\nu\lambda}\sigma_r^\lambda 
    = (\partial_\mu\bm{R}_t)^T C_{\mathrm{amb}}(\partial_\nu\bm{R}_t).
    \label{eq:appD-induced-C}
\end{equation}
The Einstein relation in the ambient space therefore gives $C_t = k_BT\gamma_t$ on $M$.

\subsubsection*{Thermodynamics defined on $M$ and comparison with the ambient thermodynamics}
We now take the Lagrange equation~\eqref{eq:appD-EQMonM}, with its restricted Lagrangian and coefficients, as the complete dynamics of a system on $M$, independently of their ambient origin.
As in Sec.~\ref{sec:appD-thermal-contact}, the heat absorbed by this system is the work done by its bath~\cite{Sekimoto_1998}.
Under the embedding $\bm{X}_t = \bm{R}_t(x_t)$ with $\bm{V}_t = \dot{x}_t^\mu \partial_\mu\bm{R}_t(x_t)$, the heat on $M$ is calculated to be
\begin{align}
    \dot{x}_t^\mu 
    \circ \left[ -(\gamma_t)_{\mu\nu}\dot{x}_t^\nu \dd{t} + \sum_r(g_t)_{\mu\nu}\sigma_r^\nu\circ\dd B_t^{(r)} \right]
    &= \dot{x}_t^\mu (\partial_\mu\bm{R}_t)^T\circ 
    \left[ - K_{\mathrm{amb}}(\partial_\nu\bm{R}_t)\dot{x}_t^\nu\dd{t} + \sum_r\bm\sigma_{\mathrm{amb},r}\circ\dd B_t^{(r)} \right]
    \nonumber\\
    &= \bm{V}_t^T \circ 
    \left[ - K_{\mathrm{amb}} \bm{V}_t \dd{t} + \sum_r\bm\sigma_{\mathrm{amb},r}\circ \dd{B}_t^{(r)} \right]
    \nonumber\\
    &= \delta Q_M,
    \label{eq:appD-heat-intrinsic}
\end{align}
where we use Eq.~\eqref{eq:appD-induced-K}.
In addition, the total entropy production rate $\dot{\Sigma}_M$ is defined by applying Eq.~\eqref{eq:appD-ambEPRDefinition}, with the same time-reversal and initial-distribution prescriptions, to the dynamics~\eqref{eq:appD-EQMonM} on $M$.
The heat is completely determined by the trajectory $\{x_{t+s}\}_{s=0}^{\Delta t}$ and changes sign under time reversal.
Therefore, this definition reduces to
\begin{equation}
 \dot{\Sigma}_M
 = \lim_{\Delta t\to0}\frac{1}{\Delta t}
 \mathbb{E}^{\mathrm F}\left[
  \log\frac{
   \mathbb{P}^{\mathrm F}[\{x_{t+s}\}_{s=0}^{\Delta t}]
  }{
   \mathbb{P}^{\mathrm R} [\{(\Theta_*x)_s\}_{s=0}^{\Delta t}]
  }
 \right]
 = \lim_{\Delta t\to 0}\frac{1}{\Delta t}
 \mathbb{E}^{\mathrm F}\!\left[
  \log\frac{
   \mathbb{P}^{\mathrm F}\bigl[\{\bm{X}_{t+s}\}_{s=0}^{\Delta t}\bigr]
  }{
   \mathbb{P}^{\mathrm R}\bigl[\{ (\Theta_*\bm{X})_{s}\}_{s=0}^{\Delta t}\bigr]
  }
 \right].
\end{equation}
Here, the second equality follows from the one-to-one correspondence between ambient trajectories and trajectories on $M$, given by $\bm{R}_{t+s}$ for the forward dynamics and $\bm{R}_{t+\Delta t-s}$ for the reverse dynamics.
The heat decomposition and Eq.~\eqref{eq:appD-ambient-entropy-split} therefore give
\begin{align}
    \delta Q_{\mathrm{amb}} &= \delta Q_M + \delta Q_b,
    \label{eq:appD-heat-comparison}
    \\
    \dot{\Sigma}_{\mathrm{amb}}
    &= \dot{\Sigma}_M - \frac{\dot Q_b}{k_BT}
    = \dot{\Sigma}_M + \frac{\mathbb{E}^{\mathrm F}[\bm b_t^TK_{\mathrm{amb}}\bm b_t]}{k_BT},
    \label{eq:appD-entropy-comparison}
\end{align}
where the mean heat rates $\dot Q_M$ and $\dot Q_b$ are defined by
$\dot{Q}_M \dd{t} \coloneqq \mathbb{E}^{\mathrm F}[\delta Q_M]$ and $\dot{Q}_b \dd{t} \coloneqq \mathbb{E}^{\mathrm F}[\delta Q_b]$.
Thus, the thermodynamics defined by the dynamics on $M$ alone accounts for the contribution determined by the particle trajectory, while the remaining contribution arises from the relative motion of the manifold and the bath.
The entropy balance defines the same system entropy rate, in units of $k_B$, whether the dynamics is described in the ambient space or on $M$:
\begin{equation}
    \dot S_{\mathrm{sys}}
    \coloneqq \dot{\Sigma}_{\mathrm{amb}} + \frac{\dot{Q}_{\mathrm{amb}}}{k_BT}
    = \dot{\Sigma}_M + \frac{\dot{Q}_M}{k_BT},
    \label{eq:appD-intrinsic-entropy-balance}
\end{equation}
where $\dot{Q}_{\mathrm{amb}} \coloneqq \dot{Q}_M + \dot{Q}_b$ is the mean ambient heat rate.

\subsubsection*{The Hamiltonian formalism and the Kramers equation}
We then derive the Kramers equation associated with the underdamped dynamics~\eqref{eq:appD-EQMonM} to take the overdamped limit.
To describe the dynamics on the phase space $T^*M\coloneqq\{(x,p)\colon x\in M,\,p\in T_x^*M\}$, we introduce the canonical momentum and Hamiltonian of the restricted Lagrangian:
\begin{align}
 p_\mu
 &\coloneqq
 \pdv{\mathcal L|_{\bm X=\bm{R}_t(x)}}{\dot x^\mu}
 =m(g_t)_{\mu\nu}\dot x^\nu
  +m(\partial_\mu\bm{R}_t)^T \partial_t\bm{R}_t,
 \nonumber\\
 H_M
 &\coloneqq
 p_\mu\dot x^\mu-\mathcal L|_{\bm X=\bm{R}_t(x)}
 \nonumber\\
 &=
 \frac{(g_t)^{\mu\nu}}{2m}
 \bigl[p_\mu-m(\partial_\mu\bm{R}_t)^T \partial_t\bm{R}_t\bigr]
 \bigl[p_\nu-m(\partial_\nu\bm{R}_t)^T \partial_t\bm{R}_t\bigr]
 +\phi_t-\frac m2|\partial_t\bm{R}_t|^2.
 \label{eq:appD-canonical-definition}
\end{align}
The canonical Hamiltonian $H_M$ generates the mechanical part of the dynamics, so that the equations of motion can be written as
\begin{align}
 \dd{x}^\mu
 &= (\nabla_{p_\mu}H_M) \dd{t},
 \nonumber\\
 \dd{p}_\mu - \Gamma^\lambda_{\mu\nu} p_\lambda\circ\dd{x}^\nu
 &=
 \Bigl[
   - \nabla_{x^\mu}H_M+f_\mu^{\mathrm{nc}} - (\gamma_t)_{\mu\nu}\nabla_{p_\nu}H_M
 \Bigr]\dd{t}
 + \sum_r [(g_t)_{\mu\nu}\sigma_r^\nu] \circ\dd B_t^{(r)},
 \label{eq:appD-canonical-dynamics}
\end{align}
where the left-hand side of the second equation is the covariant momentum increment, and $\Gamma^\lambda_{\mu\nu} \coloneqq (g_t)^{\lambda\kappa}(\partial_\kappa\bm{R}_t)^T (\partial_\mu\partial_\nu\bm{R}_t)$ is the Christoffel symbol.
The symbols $\nabla_{x^\mu}$ and $\nabla_{p_\mu}$ denote the horizontal and momentum derivatives, respectively:
\begin{equation}
    \nabla_{x^\mu} \coloneqq \nabla_\mu + \Gamma^\lambda_{\mu\nu}p_\lambda \pdv{}{p_\nu},\quad
    \nabla_{p_\mu} \coloneqq \pdv{}{p_\mu}.
\end{equation}
The horizontal derivative differentiates in position while the momentum is parallel transported~\cite[Sec.~1.1]{Lebeau2005}.
On the other hand, $H_M$ need not coincide with the physical energy $E_{\mathrm{amb}}$ for a time-dependent embedding; they satisfy
\begin{equation}
 E_{\mathrm{amb}} = H_M + m\dot{\bm X}_t^T \partial_t\bm{R}_t.
 \label{eq:appD-diff-EandH}
\end{equation}

To introduce the phase-space distribution associated with Eq.~\eqref{eq:appD-canonical-dynamics}, we write $\mathbb{E}$ for expectation over the prescribed initial-state distribution and the Wiener measure, with paths determined by that equation.
Let $\dd{\Lambda} \coloneqq (\prod_\mu\dd x^\mu)(\prod_\nu\dd p_\nu) = \dd{\mathrm{vol}_{g_t}}(x) \dd{\mathrm{vol}_{g_t^{-1}}}(p\vert x)$ be the canonical volume element on $T^*M$.
It is independent of time because the changes in the invariant position and momentum volume elements cancel~\cite[Secs.~1.1--1.2]{Lebeau2005}.
The phase-space distribution $f_M(x,p;t)$ is then defined by the relation
\begin{equation}
 \int_{T^*M} \psi(x,p) f_M(x,p;t) \dd{\Lambda} \coloneqq \mathbb{E}[\psi(x_t,p_t)]
 \label{eq:appD-phase-density}
\end{equation}
for every smooth test function $\psi$ on $T^*M$ of compact support.
The evolution equation for $f_M \dd{\Lambda}$ can be obtained by differentiating both sides with respect to time.
For the right-hand side of Eq.~\eqref{eq:appD-phase-density}, the canonical equations of motion~\eqref{eq:appD-canonical-dynamics} give
\begin{equation}
 \dv{}{t}\mathbb{E}[\psi(x_t,p_t)]
 =
    \mathbb{E}\left[
    \begin{aligned}
        &(\nabla_{p_\mu}H_M)\nabla_{x^\mu}\psi + (C_t)_{\mu\nu}\nabla_{p_\mu}\nabla_{p_\nu}\psi\\
        &-\Bigl(
        \nabla_{x^\mu}H_M-f_\mu^{\mathrm{nc}}
        +(\gamma_t)_{\mu\nu}\nabla_{p_\nu}H_M
        \Bigr)\nabla_{p_\mu}\psi
    \end{aligned}
    \right].
    \label{eq:appD-canonical-weak}
\end{equation}
Equating Eq.~\eqref{eq:appD-canonical-weak} with the time derivative of the left-hand side of Eq.~\eqref{eq:appD-phase-density}, we obtain the Kramers equation:
\begin{equation}
 \partial_t f_M
 = -\nabla_{x^\mu} \bigl[(\nabla_{p_\mu}H_M)f_M\bigr]
 +\nabla_{p_\mu}\!\left[
   \Bigl(
    \nabla_{x^\mu}H_M - f_\mu^{\mathrm{nc}} + (\gamma_t)_{\mu\nu}\nabla_{p_\nu}H_M
   \Bigr) f_M
   +(C_t)_{\mu\nu}\nabla_{p_\nu}f_M
 \right].
 \label{eq:appD-kramers}
\end{equation}

Having introduced the phase-space distribution $f_M$, we can now relate it to the entropy production.
To do so, we calculate the conditional probability ratio for a prescribed motion and its time reverse using Eq.~\eqref{eq:appD-canonical-dynamics}, as in Sec.~\ref{sec:appD-ambient-thermodynamics}.
Under the prescribed reversal, the covariant momentum increment and the mechanical force retain their signs, whereas the friction term changes sign.
For the Hamiltonian, the prescribed reversal gives $\tilde H_M(x,p;s)\coloneqq H_M(x,-p;t+\Delta t-s)$.
For an infinitesimal motion on $[t+s,t+s+\dd s]$, the corresponding reverse interval is $[\Delta t-s-\dd s,\Delta t-s]$.
Equation~\eqref{eq:appD-canonical-dynamics} gives the Gaussian bath impulses required for these two motions:
\begin{align}
    \sum_r g_{\mu\nu}\sigma_r^\nu\circ\dd B^{(r)}
    &=\dd{p}_\mu -\left[-\nabla_\mu H_M+f_\mu^{\mathrm{nc}} - \gamma_{\mu\nu}\nabla_{p_\nu}H_M\right]\dd{s},
    \\
    \sum_r \tilde{g}_{\mu\nu}\tilde{\sigma}_r^\nu \circ \dd{\tilde{B}^{(r)}}
    &=\dd{(\Theta_*p)}_\mu
    -\left[-\nabla_\mu \tilde{H}_M + \tilde{f}_\mu^{\mathrm{nc}}
    -\tilde{\gamma}_{\mu\nu}\nabla_{p_\nu} \tilde{H}_M \right] \dd{s}.
\end{align}
These Gaussian terms have the same covariance $2C\dd s$.
Therefore, the probability ratio is
\begin{align}
 \log\frac{\mathbb{P}^{\mathrm F}[x_{t+s+\dd s},p_{t+s+\dd s}\mid x_{t+s},p_{t+s}]}
 {\mathbb{P}^{\mathrm R}[x_{t+s},-p_{t+s}\mid x_{t+s+\dd s},-p_{t+s+\dd s}]}
 = -\frac{\nabla_{p_\mu}H_M}{k_BT} \circ 
 \left[\dd{p}_\mu - (-\nabla_\mu H_M + f_\mu^{\mathrm{nc}}) \dd{s} \right]
 = -\frac{\delta Q_M}{k_BT},
\end{align}
where we used the Einstein relation $C=k_BT\gamma$ and Eq.~\eqref{eq:appD-heat-intrinsic}.
Adding these log ratios over $[t,t+\Delta t]$ and including the initial-state distributions gives
\begin{equation}
 \log\frac{
 \mathbb P^{\mathrm F}[\{x_{t+s}\}_{s=0}^{\Delta t}]}
 {\mathbb P^{\mathrm R}[\{(\Theta_*x)_s\}_{s=0}^{\Delta t}]}
 =
 \log\frac{f_M(x_t,p_t;t)}
 {f_M(x_{t+\Delta t},p_{t+\Delta t};t+\Delta t)}
 -\frac1{k_BT}\int_t^{t+\Delta t}\delta Q_M.
 \label{eq:appD-path-ratio-density}
\end{equation}
Taking the expectation and the infinitesimal-interval limit, we obtain
\begin{equation}
    \dot{\Sigma}_M = - \int_{T^*M} (\partial_t f_M)\log f_M \dd{\Lambda} - \frac{\dot Q_M}{k_BT}
    = \int_{T^*M} \frac{\mathcal{J}_\mu^{\mathrm{irr}} (C_t^{-1})^{\mu\nu} \mathcal{J}_\nu^{\mathrm{irr}}}{f_M} \dd{\Lambda},
    \label{eq:appD-exact-EPR-square}
\end{equation}
where we introduce the irreversible momentum probability current $\mathcal{J}_\mu^{\mathrm{irr}} \coloneqq - (\gamma_t)_{\mu\nu}(\nabla_{p_\nu}H_M)f_M - (C_t)_{\mu\nu}\nabla_{p_\nu}f_M$.
Comparison with the entropy balance~\eqref{eq:appD-intrinsic-entropy-balance} identifies the system entropy rate as the rate of change of the Shannon entropy of $f_M$:
\begin{equation}
    \dot{S}_{\mathrm{sys}}
    = \dot{\Sigma}_M + \frac{\dot Q_M}{k_BT}
    = - \dv{}{t} \int_{T^*M} f_M \log f_M \dd{\Lambda}.
    \label{eq:appD-system-entropy-density}
\end{equation}

\subsection{Overdamped limit}
\label{sec:appD-overdamped-HM}

\subsubsection{Fokker--Planck equation}
We take $m\to0$ with the bath coefficients, forces, and
$\bm{R}_t$ held fixed~\cite{BirrellWehr2018}.
We use the small-mass expansion after the initial momentum relaxation~\cite[Sec.~3]{Matsuo_2000}.
We write $f_M(x,p;t) = \rho(x;t)P(p\vert x;t)$, where
\begin{equation}
    \rho(x;t) \coloneqq \int_{T_x^*M} f_M \dd{\mathrm{vol}_{g_t^{-1}}}(p\vert x), \quad 
    P(p\vert x;t) \coloneqq \frac{f_M}{\rho}, \quad 
    \langle A\rangle_P \coloneqq \int_{T_x^*M}AP\dd{\mathrm{vol}_{g_t^{-1}}}(p\vert x).
\end{equation}
For a smooth test function $\psi(x)$ independent of the momentum, Eq.~\eqref{eq:appD-canonical-weak} gives
\begin{equation}
 \dv{}{t}\int_M\psi\rho\dd{\mathrm{vol}_{g_t}}
 =\int_M(\nabla_\mu\psi)
 \rho\langle\nabla_{p_\mu}H_M\rangle_P
 \dd{\mathrm{vol}_{g_t}}.
\end{equation}
Thus, the position distribution $\rho$ obeys
\begin{align}
 \partial_t(\rho\dd{\mathrm{vol}_{g_t}})
 &=-\nabla_\mu(J_t)^\mu\dd{\mathrm{vol}_{g_t}},
 \nonumber\\
 (J_t)^\mu
 &\coloneqq\rho\langle\nabla_{p_\mu}H_M\rangle_P
 =\rho(g_t)^{\mu\nu}
 \left[\frac{\langle p_\nu\rangle_P}{m}
       -(\partial_\nu\bm{R}_t)^T \partial_t\bm{R}_t\right].
 \label{eq:appD-marginal-Kramers}
\end{align}
To determine the limiting current, we therefore only need $\langle p_\nu\rangle_P$ through order $m$.

Let $m^\alpha$ be the momentum scale in the small-mass limit, so that $u_\mu \coloneqq p_\mu/m^\alpha$ has an $O(1)$ scale.
The diffusion and friction terms become
\begin{align}
    \nabla_{p_\mu}\nabla_{p_\nu}[(C_t)_{\mu\nu}f_M]
    &= m^{-2\alpha} \nabla_{u_\mu}\nabla_{u_\nu}[(C_t)_{\mu\nu}f_M],
    \nonumber\\
    \nabla_{p_\mu}[(\gamma_t)_{\mu\nu}(\nabla_{p_\nu}H_M)f_M]
    &=\nabla_{u_\mu} \Bigl[ (\gamma_t)_{\mu\nu} (g_t)^{\nu\lambda}
    \bigl(m^{-1} u_\lambda - m^{-\alpha} (\partial_\lambda\bm{R}_t)^T \partial_t \bm{R}_t \bigr) f_M
    \Bigr].
\end{align}
This shows that the contribution from the embedding in the friction term is subleading to diffusion for any $\alpha>0$.
For friction and diffusion to remain in competition as $m\to0$, we must therefore have $\alpha=1/2$, so that $u_\mu = p_\mu/\sqrt{m}$.
At this scale, the remaining terms in Eq.~\eqref{eq:appD-kramers} are at most of order $m^{-1/2}$.
Solving Eq.~\eqref{eq:appD-kramers} at order $m^{-1}$ gives the Maxwell distribution as the leading term of $P$.
We therefore write
\begin{equation}
 P = P_M [1+ \sqrt{m} P^{(1)}],
 \quad
 P_M(p\vert x;t) \coloneqq \frac{\exp[-(g_t)^{\mu\nu}p_\mu p_\nu/(2mk_BT)]}{(2\pi mk_BT)^{d/2}},
 \label{eq:appD-Maxwell}
\end{equation}
where $P^{(1)}$ includes all higher corrections.

To obtain the required first moment, we substitute this expression into Eq.~\eqref{eq:appD-kramers}, multiply by $u_\alpha$, and integrate over momentum.
This gives
\begin{align}
 -\rho(C_t)_{\alpha\nu}
       \langle\nabla_{u_\nu}P^{(1)}\rangle_{P_M}
 &=\nabla_\mu\left[
       \rho\langle u_\alpha\nabla_{u_\mu}H_M\rangle_{P_M}
       \right]
 \nonumber\\
 &\quad+\rho\left[
       \langle\nabla_{x^\alpha}H_M\rangle_{P_M}
       -f_\alpha^{\mathrm{nc}}
       -(\gamma_t)_{\alpha\nu}(g_t)^{\nu\lambda}
           (\partial_\lambda\bm{R}_t)^T \partial_t\bm{R}_t
       \right]+\order{\sqrt m}.
 \label{eq:appD-correction-balance}
\end{align}
Using $\langle p_\mu\rangle_{P_M} = 0$, we can calculate the averages appearing in Eq.~\eqref{eq:appD-correction-balance} as
\begin{align}
    \langle\nabla_{u_\nu}P^{(1)}\rangle_{P_M}
    &= \frac{(g_t)^{\nu\mu}}{k_BT} \langle u_\mu P^{(1)}\rangle_{P_M}
    = \frac{(g_t)^{\nu\mu}}{mk_BT} \langle p_\mu\rangle_P,
    \nonumber\\
    \langle u_\alpha\nabla_{u_\mu}H_M\rangle_{P_M}
    &= k_BT(g_t)_{\alpha\nu} \langle\nabla_{u_\nu}\nabla_{u_\mu}H_M\rangle_{P_M}
    = k_BT\delta^\mu_\alpha,
    \nonumber\\
    \langle\nabla_{x^\alpha}H_M\rangle_{P_M} &= \partial_\alpha\phi_t + \order{\sqrt m}.
\end{align}
The last equality follows from metric compatibility, $\nabla_{x^\alpha}[(g_t)^{\mu\nu} p_\mu p_\nu] = 0$.
With the Einstein relation $C_t=k_BT\gamma_t$, Eq.~\eqref{eq:appD-correction-balance} therefore determines the current in Eq.~\eqref{eq:appD-marginal-Kramers} to the required order:
\begin{equation}
    -(\gamma_t)_{\alpha\mu} (J_t)^\mu
    = k_B T \nabla_\alpha \rho + \rho(\partial_\alpha\phi_t-f_\alpha^{\mathrm{nc}}) +\order{\sqrt m}.
    \label{eq:appD-moment-current}
\end{equation}
Taking $m\to0$, the position equation closes with
\begin{equation}
    (J_t)^\mu = -(D_t)^{\mu\nu} \left[\nabla_\nu\rho + \frac{\rho}{k_BT} (\partial_\nu\phi_t-f_\nu^{\mathrm{nc}})\right],
    \quad
    (D_t)^{\mu\nu}\coloneqq(k_BT)^2(C_t^{-1})^{\mu\nu}.
    \label{eq:appD-overdamped-current}
\end{equation}
For $f^{\mathrm{nc}} = 0$, these are Eqs.~\eqref{eq:conservation} and \eqref{eq:def_J_setup} of the main text.

\subsubsection{Heat, work, and entropy production in the overdamped limit}
We first take the overdamped limit of the mean heat, work, and entropy production accumulated over a fixed interval $[t_0,t]$, after the initial momentum relaxation, and then differentiate the limiting balances to obtain their rates.

Equation~\eqref{eq:appD-heat-intrinsic} of the intrinsic heat and Eq.~\eqref{eq:appD-canonical-weak} applied to the time-dependent test function $\psi = H_M$ give
\begin{align}
    \dot Q_M
    &=\mathbb{E}\left[ - (\gamma_t)_{\mu\nu}(\nabla_{p_\mu}H_M)(\nabla_{p_\nu}H_M) + (C_t)_{\mu\nu} \nabla_{p_\mu}\nabla_{p_\nu}H_M \right]
    \nonumber\\
    &= \dv{}{t}\mathbb{E}[H_M]- \mathbb{E}[\partial_t H_M] - \mathbb{E}[f_\mu^{\mathrm{nc}}\nabla_{p_\mu} H_M].
 \label{eq:appD-heat-weak}
\end{align}
The preceding momentum averages give, at either endpoint,
\begin{align}
    \mathbb{E}[H_M] = \frac{d}{2} k_BT + \int_M \rho\phi_t \dd{\mathrm{vol}_{g_t}} + o(1),
    \quad \mathbb{E}[E_{\mathrm{amb}}] = \mathbb{E}[H_M] + o(1),
    \label{eq:appD-endpoint-energy}
\end{align}
and, for the terms integrated in time,
\begin{align}
    \mathbb{E}[\partial_tH_M]
    &=\int_M (\partial_t\phi_t) \rho \dd{\mathrm{vol}_{g_t}} + \dot{G}_t + o(1),\qquad
    \mathbb{E}[f_\mu^{\mathrm{nc}}\nabla_{p_\mu}H_M]
    =\int_M (J_t)^\mu f_\mu^{\mathrm{nc}} \dd{\mathrm{vol}_{g_t}}
    \nonumber \\
    \dot{G}_t &\coloneqq - \frac{k_BT}{2} \int_M \rho\tr_{g_t}(\partial_t g_t) \dd{\mathrm{vol}_{g_t}}.
    \label{eq:appD-shared-energy-averages}
\end{align}
Therefore, we obtain from Eq.~\eqref{eq:appD-heat-weak}
\begin{align}
    \lim_{m\to0} \int_{t_0}^{t} \dot{Q}_M(s)\dd{s}
    &= \left[\int_M\rho\phi_s\dd{\mathrm{vol}_{g_s}}\right]_{t_0}^{t} - \int_{t_0}^{t}\left(
    \int_M \left[\rho\partial_s\phi_s + (J_s)^\mu f_\mu^{\mathrm{nc}} \right] \dd{\mathrm{vol}_{g_s}} + \dot{G}_s \right) \dd{s}
    \nonumber \\
    &= \int_{t_0}^{t} \left( \int_M \phi_s \partial_s\left( \rho \dd{\mathrm{vol}_{g_s}} \right) - \int_M (J_s)^\mu f_\mu^{\mathrm{nc}} \dd{\mathrm{vol}_{g_s}} - \dot{G}_s \right) \dd{s}
    \nonumber \\
    &= \int_{t_0}^{t} \left( \int_M (J_s)^\mu (\nabla_\mu \phi_s - f_\mu^{\mathrm{nc}} ) \dd{\mathrm{vol}_{g_s}} - \dot{G}_s \right) \dd{s}.
    \label{eq:appD-overdamped-integrated-heat}
\end{align}
The mean accumulated work in the small-mass limit is likewise calculated as
\begin{align}
    \lim_{m\to0} \int_{t_0}^{t} \dot{W}_M(s)\dd{s}
    &\coloneqq \lim_{m\to0} \int_{t_0}^{t} \left( \dv{}{s} \mathbb{E}[E_{\mathrm{amb}}] - \dot{Q}_M(s) \right) \dd{s}
    = \lim_{m\to0} \int_{t_0}^{t} \left( \dv{}{s} \mathbb{E}[H_M] - \dot{Q}_M(s) \right) \dd{s}
    \nonumber \\
    &= \int_{t_0}^{t}\left(
    \int_M \left[\rho\partial_s\phi_s+f_\mu^{\mathrm{nc}}(J_s)^\mu \right] \dd{\mathrm{vol}_{g_s}} + \dot{G}_s \right) \dd{s}.
\end{align}
Taking the time derivative of these accumulated quantities in the overdamped limit, we obtain
\begin{align}
    \dot{Q}_t^{\mathrm{od}} 
    &\coloneqq \dv{}{t}\left[\lim_{m\to0} \int_{t_0}^{t} \dot{Q}_M(s)\dd{s}\right]
    = \int_M (J_t)^\mu (\nabla_\mu \phi_t - f_\mu^{\mathrm{nc}} ) \dd{\mathrm{vol}_{g_t}} - \dot{G}_t, \\
    \dot{W}_t^{\mathrm{od}} 
    &\coloneqq \dv{}{t}\left[\lim_{m\to0} \int_{t_0}^{t} \dot{W}_M(s)\dd{s}\right]
    = \int_M \left[ (\partial_t\phi_t) \rho + (J_t)^\mu f_\mu^{\mathrm{nc}} \right] \dd{\mathrm{vol}_{g_t}} + \dot{G}_t.
\end{align}
For $f_\mu^{\mathrm{nc}}=0$, these reproduce Eqs.~\eqref{eq:Q_od_rate} and \eqref{eq:W_od_rate} in the main text.

The total entropy production rate and the rate of change of the system entropy are obtained by substituting $f_M=\rho P_M(1+\order{\sqrt m})$ into Eq.~\eqref{eq:appD-system-entropy-density} and following the same procedure:
\begin{align}
    \dot S_{\mathrm{sys}}^{\mathrm{od}}
    &\coloneqq \dv{}{t}\left[ \lim_{m\to0}\int_{t_0}^{t}\dot S_{\mathrm{sys}}(s) \dd{s} \right]
    = \dv{}{t} \left( - \int_M \rho \log \rho \dd{\mathrm{vol}_{g_t}} \right),
    \nonumber \\
    \dot{\Sigma}_t^{\mathrm{od}}
    &\coloneqq \dv{}{t} \left[ \lim_{m\to0} \int_{t_0}^{t}\dot{\Sigma}_M(s)\dd{s} \right]
    = \dot S_{\mathrm{sys}}^{\mathrm{od}} - \frac{\dot{Q}_t^{\mathrm{od}}}{k_B T}
    = \int_M \frac{(J_t)^\mu(D_t^{-1})_{\mu\nu}(J_t)^\nu}{\rho} \dd{\mathrm{vol}_{g_t}} 
    \ge 0.
    \label{eq:appD-overdamped-EPR}
\end{align}
For $f_\mu^{\mathrm{nc}}=0$, these expressions reproduce the entropy production rate in Eq.~\eqref{eq:diff_RelativeEntropy} and the entropy balance in Eqs.~\eqref{eq:EP_total}--\eqref{eq:def_S_sys} of the main text.
\par
\endgroup

\begingroup
\section{The Fokker--Planck equation as a gradient flow of the relative entropy}
\label{sec:supp_gradient_flow}
\endgroup

In this section, we show that the {dynamics defined by Eqs.~(\ref{eq:conservation}) and (\ref{eq:def_J_setup})} in the main text can be written as a gradient flow of the relative entropy with respect to the instantaneous equilibrium measure ${\mu_t^{\rm eq}}$. {In this section, we assume that $M$ is compact.}\par
We begin by giving the definition of a gradient flow. {Consider a Riemannian manifold $(\mathcal{N},\mathcal{G})$.} {For a scalar function \(f\in C^\infty(\mathcal{N})\), we define the gradient \(\operatorname{grad}_{\mathcal{G}}f\) with respect to the metric $\mathcal{G}$ by requiring, for all vector fields \(X\),
\begin{align}
    \mathcal{G}(\operatorname{grad}_{\mathcal{G}}f,X)=\dd f(X),
\end{align}
where $\mathcal{G}(X,Y)=\mathcal{G}_{\mu\nu}X^\mu Y^\nu$ and $\dd f(X)
  =
  X^\mu \partial_{x^\mu}f$, or equivalently
\begin{align}
    (\operatorname{grad}_{\mathcal{G}}f)^\mu
  =
  \mathcal{G}^{\mu\nu}\partial_{x^\nu}f .
\end{align}
}
For a smooth function $\mathcal E:\mathcal N\to\mathbb R$ on a Riemannian manifold 
$(\mathcal N,\mathcal G)$ {and a time interval $[0,\tau]$}, a curve $z:[0,\tau]\to \mathcal{N}$ is called the negative gradient flow of $\mathcal E$ if
\begin{align}
  \frac{\dd z(s)}{\dd s}
  =
  -
  \operatorname{grad}_{\mathcal G}\mathcal E(z(s)),
\end{align}
that is, at each time $s$, the curve evolves in the direction of steepest descent of 
$\mathcal{E}(z(s))$ from the point $z(s)$.

We now show that {Eqs.~(\ref{eq:conservation}) and (\ref{eq:def_J_setup})} can be written as a gradient flow {on $\mathcal P^\infty(M)$}. We fix $t$, $g_t$, $\phi_t$, and $(D_t)^{\mu\nu}$. We define
\begin{align}
  \mathcal P^\infty(M)
  \coloneqq
  \left\{
    \mu=\rho_t\,\dvolg
    \ \middle|\
    \rho_t\in C^\infty(M),\
    \rho_t>0,\
    \int_M\rho_t\,\dvolg=1
  \right\},
\end{align}
and introduce a generalized Wasserstein metric on $\mathcal P^\infty(M)$.
For a tangent vector $\xi=\sigma\,\dvolg\in T_\mu{\mathcal P^\infty(M)}$ at
$\mu=\rho_t\,\dvolg\in{\mathcal P^\infty(M)}$, where the tangent space is {formally given by}
\begin{align}
  T_\mu{\mathcal P^\infty(M)}
  \coloneqq
  \left\{
    \xi=\sigma\,\dvolg
    \ \middle|\
    \sigma\in C^\infty(M),\
    \int_M\sigma\,\dvolg=0
  \right\},
\end{align}
we define the potential $\psi_\xi\in C^\infty(M)$ as the function satisfying
\begin{align}
  -
  \nabla_\mu
  \left[
    \rho_t (D_t)^{\mu\nu}
    \nabla_\nu\psi_\xi
  \right]
  =
  \sigma,
  \label{eq:supp_potential_equation}
\end{align}
which is unique up to an additive constant. The generalized Wasserstein metric $\mathcal{W}_{t,\mu}$ {on $\mathcal P^\infty(M)$} is then defined for $\xi,\eta\in T_\mu{\mathcal P^\infty(M)}$ by
\begin{align}
  \mathcal W_{t,\mu}(\xi,\eta)
  \coloneqq
  \int_M
  \rho_t\,
  (D_t)^{\mu\nu}
  \left[
    \nabla_\mu\psi_\xi
  \right]
  \left[
    \nabla_\nu\psi_\eta
  \right]
  \dvolg,
  \label{eq:supp_W_metric}
\end{align}
which is a generalized Wasserstein metric with the diffusion tensor $(D_t)^{\mu\nu}$ as the mobility~\cite{Lisini_2008}.\par
We next compute the gradient of the relative entropy functional
\begin{align}
  \mathcal S_t(\mu)
  \coloneqq
  S(\mu\|{\mu_t^{\rm eq}})
  =
  \int_M
  \rho_t
  \ln\frac{\rho_t}{\rho_t^{\rm eq}}
  \dvolg
\end{align}
with respect to this metric.
For any tangent vector $\xi=\sigma\dvolg\in T_\mu{\mathcal P^\infty(M)}$,
{let $\dd\mathcal S_{t,\mu}[\xi]$ denote the directional derivative of
$\mathcal S_t$ at $\mu$ along $\xi$. Using} $\int_M\sigma\dvolg=0$, we have
\begin{align}
  \dd\mathcal S_{t,\mu}[\xi]
  &=
  \left.
  \frac{\dd}{\dd\epsilon}
  \int_M
  (\rho_t+\epsilon\sigma)
  \ln
  \frac{\rho_t+\epsilon\sigma}{\rho_t^{\rm eq}}
  \dvolg
  \right|_{\epsilon=0}
  \notag\\
  &=
  \int_M
  \ln\frac{\rho_t}{\rho_t^{\rm eq}}\,
  \sigma\,\dvolg \notag\\
  &=
  -\int_M
  \ln\frac{\rho_t}{\rho_t^{\rm eq}}\,
  \nabla_\mu
  \left[
    \rho_t (D_t)^{\mu\nu}
    \nabla_\nu\psi_\xi
  \right]
  \dvolg
  \notag\\
  &=
  \int_M
  \rho_t (D_t)^{\mu\nu}
  \left[
    \nabla_\mu
    \ln\frac{\rho_t}{\rho_t^{\rm eq}}
  \right]
  \left[
    \nabla_\nu\psi_\xi
  \right]
  \dvolg
  \notag\\
  &=
  \mathcal W_{t,\mu}
  \left(
    -
    \nabla_\alpha
    \left[
      \rho_t (D_t)^{\alpha\beta}
      \nabla_\beta
      \ln\frac{\rho_t}{\rho_t^{\rm eq}}
    \right]\dvolg,
    \xi
  \right).
\end{align}
By the definition of the gradient, we therefore obtain
\begin{align}
  \operatorname{grad}_{\mathcal W_t}\mathcal S_t(\mu)
  =
  -
  \nabla_\mu
  \left[
    \rho_t (D_t)^{\mu\nu}
    \nabla_\nu
    \ln\frac{\rho_t}{\rho_t^{\rm eq}}
  \right]\dvolg.
  \label{eq:supp_grad_entropy}
\end{align}
The negative gradient flow of $\mathcal S_t$ is thus
\begin{align}
  \partial_s(\rho_s\dvolg)
  &=
  -
  \operatorname{grad}_{\mathcal W_t}\mathcal S_t(\mu_s)
  \notag\\
  &=
  \nabla_\mu
  \left[
    \rho_s(D_t)^{\mu\nu}
    \nabla_\nu
    \ln\frac{\rho_s}{\rho_t^{\rm eq}}
  \right]\dvolg,
  \label{eq:supp_negative_gradient_flow}
\end{align}
which coincides with the {dynamics defined by Eqs.~(\ref{eq:conservation}) and (\ref{eq:def_J_setup})} in the main text.

\addvspace{\baselineskip}
\section{A colloidal particle on a sphere with a time-evolving radius}
\label{sec:supp_sphere}
\begingroup
As an additional example, we consider a colloidal particle on a sphere whose radius evolves in time. Such a setup {may} be realized experimentally by varying the radius of a giant vesicle via osmotic pressure~\cite{Viallat_2004} and controlling adsorbed particles with optical traps~\cite{Dimova_2000}. Let $M$ be a {2-dimensional} sphere of radius $r_t$ with
{angular coordinates $(\theta,\phi)$ and} metric $(g_t)_{\theta\theta}={r_t}^2$, $(g_t)_{\varphi\varphi}={r_t}^2\sin^2\theta$, $(g_t)_{\theta\varphi}=0$. A straightforward calculation yields
\begin{align}
\dot{G}_t&=-2\kB T\frac{\dot r_t}{r_t}.
\label{eq:God_sphere_rate}
\end{align}
Substituting this into Eqs.~(\ref{eq:Q_od_rate}) and (\ref{eq:W_od_rate}) shows that the particle receives negative work and absorbs heat when the sphere expands, and vice versa when it contracts. This follows from the scaling $E_t^\mathrm{kin}\propto {r_t}^{-2}$ at fixed components of conjugate momentum $p_\theta=m{r_t}^2\dot\theta$ and $p_\varphi=m{r_t}^2\sin^2\theta\,\dot\varphi$, which makes the kinetic energy change oppositely to the radius. Since the work done by the metric equals the change in kinetic energy, its sign is opposite to that of the radius change.\par
We use this example to numerically verify the expressions for heat and work in Eqs.~(\ref{eq:Q_od_rate}) and (\ref{eq:W_od_rate}). {For the numerical calculation, we take isotropic friction, $(\gamma_t)_{\mu\nu}=\gamma(g_t)_{\mu\nu}$, with a constant $\gamma>0$.} The radius of the sphere is contracted from $r_0$ to {$r_\tau\ (<r_0)$} over the time interval 
$t\in[0,\tau]$ according to
\begin{align}
  r_t=r_0+(r_\tau-r_0)\frac{t}{\tau}.
\end{align}
We consider the transport of a particle under the potential
\begin{align}
  \phi_t(x)
  &=
  \kappa {r_t}^2
  \left[
    1-\sqrt{1-\cos^2\theta}\,
    \cos\left(\varphi-\pi \frac{t}{\tau}\right)
  \right],
\end{align}
whose minimum traces a half-circle along the equator. We set the initial distribution to be $\rho_0(x)\propto \exp[-\phi_0(x)/(k_\B T)]$.\par
\begin{figure}[!htb]
  \centering
  {\includegraphics[width=\dimexpr(\textwidth-\columnsep)/2\relax]{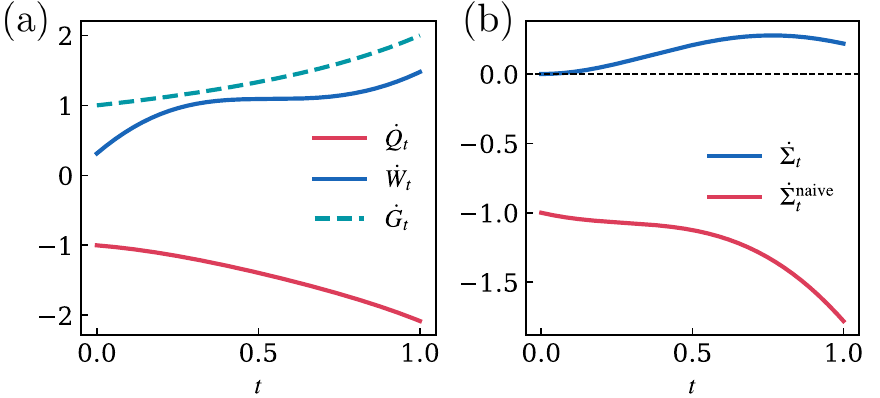}}
  \caption{
(a) Time evolution of $\dot{Q}_t$, $\dot{W}_t$, and $\dot{G}_t$ obtained from the numerical simulation. The parameters are $k_{\mathrm{B}}T=1.0$, $\gamma=5.0$, $\tau=1.0$, $r_0=1.0$, $r_\tau=0.5$, and $\kappa=1.0$.
(b) Comparison of $\dot{\Sigma}_t$ and $\dot{\Sigma}_t^{\mathrm{naive}}$.
The correct entropy production rate $\dot{\Sigma}_t$ remains non-negative, whereas the naive one $\dot{\Sigma}_t^{\mathrm{naive}}$ can become negative.
}
  \label{fig:numresults}
\end{figure}

We numerically solve the overdamped Fokker--Planck dynamics defined by Eq.~(\ref{eq:conservation}) and the current~(\ref{eq:def_J_setup}) in this setup to verify the main claim. Figure~\ref{fig:numresults}(a) shows the time evolution of {$\dot{Q}_t$, $\dot{W}_t$, and $\dot{G}_t$}. As discussed above, {$\dot{G}_t$} appears as a positive contribution that corrects both heat and work. Figure~\ref{fig:numresults}(b) compares {$\dot{\Sigma}_t$ with $\dot\Sigma_t^{\mathrm{naive}}$, demonstrating that the correct entropy production rate $\dot\Sigma_t$ is consistent with the second law, whereas the naive one $\dot\Sigma_t^{\mathrm{naive}}$ yields negative values, violating the second law.}
\par\endgroup

\makeatletter
\begingroup
\renewcommand{\bibsection}{\par\addvspace{\baselineskip}\section*{Supplemental references}}
\renewenvironment{thebibliography}[1]
  {\NAT@thebibliography{#1}\setcounter{NAT@ctr}{\letterreferencecount}}
  {\endNAT@thebibliography}
\endgroup
\makeatother

\end{document}